\documentclass[useAMS,usenatbib,refree]{mn2e}
\usepackage{graphicx,amsmath,amssymb,color}

\newcommand{\betacr}{ \beta_{\rm c}} 
\newcommand{\betastar}{ \beta_{\ast} } 
\newcommand{\tevolve}{ t_{\rm evolve}} 
\newcommand{\rhobase}{ \rho_{\rm b}} 
\newcommand{\hscale}{ L_{\rm s}} 
\newcommand{\rout}{ R_{\rm out} } 
\newcommand{\rplan}{ R_{\rm p} } 
\newcommand{\mplan}{ M_{\rm p} } 
\newcommand{\bplan}{ B_{\rm p} } 
\newcommand{\cs}{ c_{\rm s} } 
\newcommand{\bc}{}

\title[Hot Jupiter Breezes]{Hot Jupiter Breezes: \\
Time-dependent Outflows from Extrasolar Planets}
\author[J. E. Owen \& F. C. Adams]
{James E. Owen$^{1}$\thanks{E-mail: jowen@ias.edu}\thanks{Hubble Fellow} and Fred C. Adams$^{2,3}$\\
$^{1}$Institute for Advanced Study, Einstein Drive, Princeton NJ, 08540, USA\\
$^{2}$Physics Department, University of Michigan, Ann Arbor, MI 48109, USA \\
$^{3}$Astronomy Department, University of Michigan, Ann Arbor, MI 48109, USA }

\begin{document}
\include{journals_mnras}

\maketitle

\label{firstpage}

\begin{abstract}
We explore the dynamics of magnetically controlled outflows
from Hot Jupiters, where these flows are driven by UV heating from the
central star.  In these systems, some of the open field lines do not
allow the flow to pass smoothly through the sonic point, so that
steady-state solutions do not exist in general. This paper focuses on
this type of magnetic field configuration, where the resulting flow
becomes manifestly time-dependent. We consider the case of both steady
heating and time-variable heating, and find the time scales for the
corresponding time variations of the outflow. Because the flow cannot
pass through the sonic transition, it remains subsonic and leads to
so-called breeze solutions. One manifestation of the time variability
is that the flow samples a collection of different breeze solutions
over time, and the mass outflow rate varies in quasi-periodic fashion.
Because the flow is subsonic, information can propagate inward from
the outer boundary, which determines, in part, the time scale of the
flow variability. This work finds the relationship between the outer
boundary scale and the time scale of flow variations.  In practice,
the location of the outer boundary is set by the extent of the sphere
of influence of the planet. The measured time variability can be used,
in principle, to constrain the parameters of the system (e.g., the 
strengths of the surface magnetic fields). 
\end{abstract}


\section{Introduction}
\label{sec:intro} 

Extrasolar planets that reside sufficiently close to their host stars
can experience evaporation driven by intense stellar heating,
primarily from radiation at ultraviolet (UV) wavelengths. For planets
with smaller masses, roughly $\mplan<0.1M_J$, planetary masses can be
substantially reduced through the photoevaporative process \citep{owenwu}. 
For planets with larger masses, roughly $\mplan>0.5M_J$, mass loss
rates are suppressed, and total planetary masses are not significantly
altered. In these higher mass cases, however, the photoevaporative
flows can be observed and provide us with important insights regarding
the interactions between stars and close planets. Thus, these outflows
represent a promising channel through which we can characterize the
atmospheres of exoplanets. In spite of its potential importance, only
a modest amount of theoretical work has been carried out to date (see
the discussion below). The goal of this paper is to generalize our
understanding of the process of planetary evaporation, with a focus on
the roles played by magnetic fields and time-varying flows. More
specifically, some magnetic field configurations do not allow the
outflow to pass smoothly through a sonic transition, so that the flow
must display time variations.

Evidence for the evaporation of Hot Jupiters has been observed for at 
least two extrasolar planetary systems, including HD 209458
\citep{vidalmadjar} and HD 189733 \citep{lecavelier}, {\bc with the later exhibiting variability in Lyman-$\alpha$ measurements from transit to transit \citep{lecavelier2012}. Additionally, an outflow from a close-in Neptune mass planet has been recently reported \citep{Ehrenreich2015}.} 
The mass loss rates for these Hot Jupiters are inferred to be of order
${\dot M} \approx 10^{10} - 10^{11}$ g s$^{-1}$, which is roughly
equivalent to $10^{-3} M_J$ Gyr$^{-1}$. Moreover, these mass loss
rates are comparable (in order of magnitude) to those expected
theoretically from simple energetic considerations, where a
substantial fraction of the incoming UV flux is converted into the
mechanical luminosity of the outflow (e.g., \citealt{watson,lammer}). {\bc The ``efficiency'' of the heating process has recently been estimated for extreme ultraviolet heating of Hydrogen dominated atmospheres, indicating it is $\lesssim 20\%$ \citep{shematovich2014}}. 
On a related front, observational evidence for star-planet
interactions has also been reported, and the relevant signatures 
vary substantially from epoch to epoch \citep{shkolnik2005,shkolnik2008}.

With the observational discoveries outlined above, several theoretical
treatments of mass loss from planetary surfaces have been developed.
The first set of calculations considered simple spherical flows
\citep{watson,lammer,baraffe}. This work was then generalized to
include chemistry, photoionization, recombination, tidal potentials,
X-ray heating, and two-dimensional geometry {\bc (e.g., see
\citealt{yelle,lecavelier2004,munoz,Murray-Clay2009a,stoneproga,Owen2012c,koskinen2013a,koskinen2013b,oa15}} and
references therein).

Significantly, the {\bc majority} of aforementioned treatments of planetary outflows do
not include magnetic fields, {\bc \citep[see][for initial treatments]{Adams2011,trammell2011,Bisikalo2013,Owen2014f,Bisikalo2015}}. However, the giant planets in our solar
system, as well as most stars, have internal magnetic fields, and we
might expect hot Jupiters to support fields of comparable strength.
For a given mass loss rate, we can estimate the importance of magnetic
fields by determining the ratio of the ram pressure of the outflow to
the magnetic field pressure. This dimensionless quantity (see also 
\citealt{Owen2014f}) can be written in the form 
\begin{equation}
\Lambda = \frac{8\pi\rho v^2 }{ B^2} = \frac{2 {\dot M} v }{ B^2 r^2} \,,
\end{equation}
which is a function of position, and 
which can be evaluated using typical values to obtain 
\begin{eqnarray}
\Lambda &\approx& 2 \times 10^{-4} 
\left( \frac{{\dot M} }{ 10^{10}\,\,{\rm g} \,\,{\rm s}^{-1}} \right) 
\left( \frac{v }{ 10 {\rm km}\,\,{\rm s}^{-1}} \right) \nonumber \\
&\times&\left( \frac{B }{ 1 {\rm G}} \right)^{-2} 
\left( \frac{r }{ 10^{10}\,\,{\rm cm}} \right)^{-2} \,. 
\label{magratio} 
\end{eqnarray}
For a surface field strength of 1 gauss, the magnetic field pressure
is thus larger than the ram pressure of the outflow by a factor of
$\sim10^4$ at the planetary surface, and this ratio increases to
$\sim10^6$ at the sonic point (see also \citealt{Adams2011,Owen2014f}). 
As a result, the flow must be magnetically controlled: The outflow is
not strong enough to bend the magnetic field lines and must instead
follow their geometry (which is set by independent physical
processes).

One-dimensional, steady-state models of magnetically controlled
outflows from planets have been developed
\citep{Adams2011,trammell2011}, and this work has recently been
generalized to two dimensions using numerical simulations
\citep{Owen2014f,trammell2014}. Unlike the case of spherically
symmetric flow \citep{Parker1958,Shu1992}, however, the conditions
required for the flow to pass smoothly through the (generalized) sonic
point are non-trivial \citep{Adams2011}. In particular, no smooth
transitions exist for flow that follows some of the open field
lines, specifically those with small divergence. Under such circumstances, no steady-state solutions are
available and the flow must display time-dependent behaviour, as seen
in numerical simulations \citep{Owen2014f}.  Moreover, when the flow
cannot smoothly transition from subsonic to supersonic conditions, the
outflows can remain subsonic, and such cases are often labelled as
``breeze solutions''. However, the breeze solutions require a large
finite pressure at infinity in order to be the time-independent. The
large required pressures are inconsistent with those found in the
environment around the planet, so that we expect these breeze
solutions to be unstable on dynamic time-scales (see also the
arguments presented in \citealt{Parker1958} with applications to the
solar wind). The objective of this present work is thus to understand
magnetically controlled breeze solutions from planetary surfaces.
These flows are driven by the UV heating from the central stars and
are necessarily time-dependent.

This paper is organized as follows. The basic problem is formulated in
Section \ref{sec:formulate}, where we develop the equations of motion
for flow along a given magnetic field line. This treatment uses a
coordinate system where one coordinate follows the field direction, so
that the coordinate system depends on the underlying magnetic field
configuration (including the background contribution from the star).
We then describe the numerical methods and the boundary conditions
(see also the Appendices).  In Section \ref{sec:nonsteady}, we present
outflow solutions for configurations that support steady flow and
those that do not, with a focus on the transition between the two
cases.  The generalization to non-steady forcing functions is
presented in Section \ref{sec:driven}, where we explore the
relationship between the driving time scale and the corresponding time
variations in the outflows. The implications of this work are
considered in Section \ref{sec:imp}, with a focus on possible
observational signatures, including an overview of the relevant time
scales in the problem.  Finally, the paper concludes in Section
\ref{sec:conclude} with a summary of results and further discussion of
their possible applications. 

\section{Formulation of the Problem}
\label{sec:formulate} 

Since the flow from Hot Jupiters is likely to be strongly magnetically
controlled \citep{Owen2014f}, we choose to analyse the problem on a
streamline by streamline basis. In this case, the flow problem can be
reduced to a one-dimensional calculation along the geometry of a
single magnetic field line. Following previous work
\citep{Adams2011,Adams2012}, we define a set of orthogonal coordinates
that follow an axi-symmetric magnetic field. Using this set of
coordinates one can reduce the multi-dimensional MHD problem down to a
simple set of 1D calculations along each field line.

In this work we take a simple but plausible form for the magnetic
field configuration of the star/planet system, but note that this
approach can be readily generalised to any axi-symmetric field.
Specifically, we take a magnetic field topology of the following form:
The field has a dipole contribution from the planet that points in a
direction that is perpendicular to the planet's orbital plane and a
constant background component from the star that points in the same
direction as the dipole.\footnote{Note that the stellar background
  field points in the same direction as the planetary dipole when the
  stellar dipole is anti-aligned with that of the planet.} For this
choice, the magnetic field has the following configuration: 
\begin{equation}
{\bf B}=\bplan\left[\xi^{-3}\left(\cos\theta{\bf\hat{r}}-
{\bf{\hat{z}}}\right)+\betastar{\bf{\hat{z}}}\right]\,,
\end{equation}
where $\bplan$ is the magnetic field strength at the planet surface,
$\xi$ is a dimensionless radial coordinate, measured in terms of the
planet's radius ($\rplan$), and $\betastar$ is a parameter that
measures the background field strength of the star and is defined by: 
\begin{equation}
\betastar=\frac{B_*}{\bplan}\left(\frac{R_*}{a}\right)^3\,,
\end{equation}
where $a$ is the orbital separation. Given this magnetic field
topology, an orthogonal coordinate system $\{p,q,\phi\}$ can be
constructed from the standard spherical polar coordinate system
$\{\xi,\theta,\phi\}$, such that: 
\begin{eqnarray}
p&=&\left(\betastar\xi-\xi^{-2}\right)\cos\theta\\
q&=&\left(\betastar\xi^{2}+2\xi^{-1}\right)^{1/2}\sin\theta
\end{eqnarray}
and the coordinate system scale factors $\{h_p,h_q,h_\phi\}$ can be
constructed trivially (see \citealt{Adams2011}). We note that along a
given field line $q$ is a constant and introduce $\theta_0$ as the
polar angle the field line has at the planet's surface, which is
constant a given field line and more physically meaningful than
labelling field lines by $q$.

For simplicity we assume the flow is isothermal with sound speed
($\cs$) and an ideal equation of state, such that the pressure is
given by $P=\cs^2\rho$. An isothermal flow is a good approximation for
the high fluxes expected at early times, when the flow is in approximate
local thermodynamic equilibrium at $T\sim 10^{4}$~K
\citep{Murray-Clay2009a,Owen2012c,Owen2014f}. Therefore, the
one-dimensional equations of hydrodynamics along a field line become:
\begin{eqnarray}
\frac{\partial \rho}{\partial t}&=&-\frac{1}{h_ph_qh_\phi}
\frac{\partial}{\partial p}\left(h_qh_\phi\rho u\right)\label{eqn:rho} \\
\frac{\partial u}{\partial t}+\frac{u}{h_p}\frac{\partial u}{\partial p}&
=&-\frac{1}{h_p\rho}\frac{\partial P}{\partial p}-\frac{1}{h_p}
\frac{\partial \Psi}{\partial p} \,.\label{eqn:u}
\end{eqnarray} 
Finally, we work in dimensionless units, such that $G=\cs=\rplan=1$, although we will convert back to dimensional parameters in the discussion.
This choice implies that the time variable for the simulation is in
units of the sound crossing time of the planet $t=\rplan/\cs$, which has
a value $\sim10^4$~s for a Hot Jupiter with radius $10^{10}~$cm. The
density is scaled such that the density of the outflow at the
planetary surface is unity. When we consider cases where the base
density of the flow varies in time, the lowest amplitude of the
variation is chosen to have a density of unity. In this formalism, 
the dimensionless potential becomes
\begin{equation}
\psi=-\frac{b}{\xi}\,,
\end{equation}
where $b=G\mplan/(\cs^2\rplan)$ measures the depth of the gravitational
potential of the planet. Note that, for simplicity, we ignore the
contribution from the tidal potential due to the star-planet
interaction. For a typical hot Jupiter with $\mplan=0.8M_J$ and
$\rplan=10^{10}$~cm, the dimensionless depth of the potential
$b\approx10$. Equations~(\ref{eqn:rho}) and (\ref{eqn:u}) represent
the time-dependant flow along a given field line. For open field
lines, steady-state trans-sonic flow solutions that pass smoothly
through the sonic point exist \citep{Adams2011}, provided that the
parameter $\betastar$ falls below a critical value. We also note that
in the absence of rotation, a non-zero value of $\betastar$ is
required to open up the field lines and thereby allow for 
outflow.\footnote{This statement holds for the case of magnetically 
controlled flow. If the ram pressure is sufficiently high, the outflow 
itself can open up field lines.} 

\subsection{Numerical Method}

We build a {\sc zeus}-style one-dimensional hydrodynamics code
\citep{Stone1992,Hayes2006} that solves our problem (e.g.,
Equations~[\ref{eqn:rho}] and [\ref{eqn:u}]). We make use of operator
splitting and split the hydrodynamic update into three sub-steps
\citep{Stone1992}. First we update the velocity due to the force terms: 
\begin{equation}
\frac{\partial u}{\partial t}=-\frac{1}{h_p\rho}
\frac{\partial P}{\partial p}-
\frac{1}{h_p}\frac{\partial \Psi}{\partial p}\,.
\end{equation}
Next the velocity is updated due to the artificial viscosity
\begin{equation}
\frac{\partial u}{\partial t}=-\frac{\left(\nabla \cdot {\bf Q}\right)_p}{\rho}\,,
\end{equation}
where ${\bf Q}$ is the artificial viscosity tensor. However, unlike
{\sc zeus} we find it necessary to include a full implementation of
$\nabla\cdot{\bf Q}$, {\sc zeus} uses a von Neumann \& Richtmyer approach
\citep{VonNeumann1950} which neglects the curvature terms. The
expressions for this in our coordinate system and detailed in
Appendix~\ref{sec:TV}. Finally, we perform an advection update of the
form: 
\begin{eqnarray}
\frac{\rm d}{{\rm d}t}\int_V\!\!\!{\rm d}V\,\rho&=
&-\int_{\partial V}\!\!\!\!\rho{\bf u}\cdot{\rm d}{\bf S} \\
\frac{\rm d}{{\rm d}t}\int_V\!\!\!{\rm d}V\,\rho{\bf u}&=
&-\int_{\partial V}\!\!\!\!\rho{\bf uu}\cdot{\rm d}{\bf S}
\end{eqnarray}
The equations are updated on a staggered mesh, where scalars
(e.g., $\rho$) are stored at cell centres and vectors (e.g., $u$) are
stored at cell boundaries. For the advection step our upwinded fluxes
are reconstructed using a second order method with a van-Leer limiter
\citep{VanLeer1977}. The explicit time-step is subject to a CFL
condition and additionally constrained such that it cannot increase by
more than 30\% each step. The artificial viscosity constant is set
such that any discontinuities are spread over three cells
(i.e., $C_v=3$ -- see Appendix~\ref{sec:TV}). The tests we have
performed to ensure our code is behaving as expected are detailed in
Appendix~\ref{sec:code_tests}.

Since, we are expecting --- and indeed find --- time variable outflow,
we consider two scenarios. We first consider the case where the system
has no sources of time-variability at any of the boundaries, so that
any variability arises from the physics of the flow itself, e.g., the
absence of smooth sonic transitions. This approach is discussed in
Section \ref{sec:nonsteady}.  In addition, the high energy radiative
output of the star is likely to be time-variable, e.g., due to stellar
flares. We thus consider `driven' flows, where we vary the density at
the base of the wind on a characteristic time-scale; this version of
the problem is discussed in Section \ref{sec:driven}.

\begin{figure*}
\centering
\includegraphics[width=\textwidth]{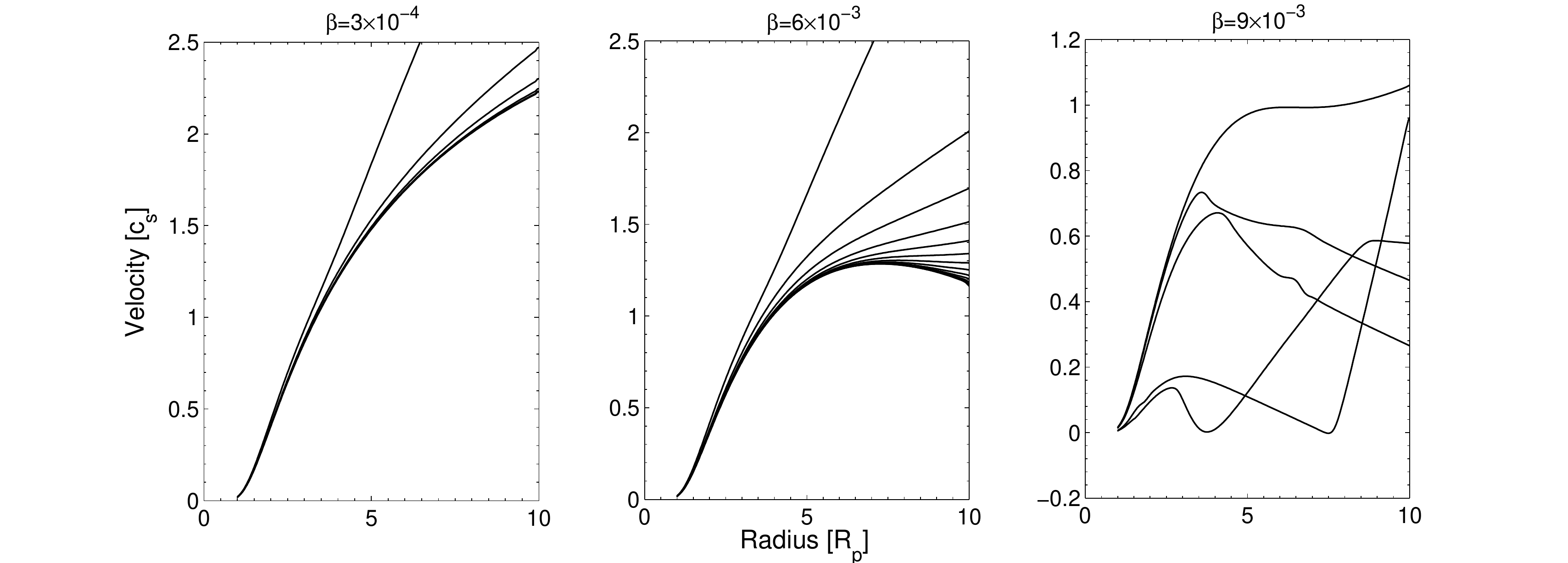}
\caption{Snapshots of the flow velocity structure as a function of  
(spherical) radius from the planet for a system with $b=10$ and  
$\theta_0=0.1$. The three panels show results for magnetic field  
parameter $\betastar=3\times10^{-4}$, $6\times10^{-3}$, and
$9\times10^{-3}$. For the two low $\betastar<\betacr$ cases (left and
middle panel), the snapshots are plotted every $t=3$ up to a total
time of 30. For the case with high $\betastar>\betacr$ (right panel), 
the snapshots are plotted every $t=20$ up to a total time of $t=100$.
Note that this latter case does not converge to a smooth, steady-state
solution (see text). } 
\label{fig:crit_beta}
\end{figure*}  

\subsection{Choice of Boundary Condition}

One part of the advantage of our approach is we have much more control
over the implementation of the boundary conditions than in the
multi-dimensional calculations \citep{Owen2014f}. At the inner
boundary the choice depends on whether the gas velocity on the
boundary is inwards or outwards. If the velocity is outwards the
choice is easy and the ghost cells are assigned the base density
hydrostatic structure $\rhobase$ and the velocity structure is chosen to
feed the domain with the required mass-flux. If the gas velocity is
inwards we have two choices: either implement outflow boundary
conditions or zero mass flux boundary conditions. The former would be
appropriate if any inward flow could disturb the underlying atmosphere
and the latter if it could not. Since the multi-dimensional
calculations which included a portion of the unheated atmosphere in
the simulation domain show that any inflow that occurred was sub-sonic
and therefore did not disturb the atmosphere then a zero-mass flux
boundary is the more physical choice. Therefore, if we detect inflow
at the inner boundary the velocity is set to zero and we retain the
hydrostatic density structure in the ghost zones.

For the outer boundary again the choice of boundary condition depends
on whether the gas velocity is inflowing our outflowing. In the case
of outflow we use outflow boundary conditions. However, due to the
fact the outflow velocity is often sub-sonic standard extrapolations
techniques employed often lead to spurious reflections from the outer
boundary, which propagate back into the simulation domain and lead to
non-physical flow structure developing. Instead we apply
characteristic tracing boundary conditions \citep{Thompson1987}, where
the values in the ghost zones are computed assuming that only outgoing
characteristics at the boundary and any incoming characteristics are
given zero amplitude. This prevents any spurious reflections and the
details of our implementation is detailed in the Appendix. In the case
we detect inflow at the outer boundary we set the velocity to zero and
the density to the ambient value due to the presence of the star's
wind/atmosphere (i.e., $\rho=10^{-6}\rhobase$), although we note that the
choice of this value has little effect on the dynamics unless it is
comparable to the hydrostatic value.

\section{Transition to Non-steady Flow and Time Variability}
\label{sec:nonsteady} 

For a given magnetic field line, there exists a critical value of the
background stellar field (specified as $\betacr$), above which no
steady-state solutions exist that satisfy the required boundary
conditions (i.e., $\rho\rightarrow 0$ as $r\rightarrow\infty$). This
claim includes hydrostatic solutions. As a result, the only possible
outflows must display non-steady (time-varying) flow. Moreover, the
critical value of the magnetic field strength ratio ($\betacr$) is
predicted by analytic theory \citep{Adams2011}. To demonstrate this
transition, from systems that allow steady-state flow to those that do
not, we construct a set of simulations where we increase $\betastar$
from below $\betacr$ to above.

For high launching latitudes, the critical, $\betastar$ is given by
$\betacr\approx8/b^3$ \citep{Adams2011}. For our ``standard'' set of
simulation parameters, we choose $b=10$ and $\theta_s=0.1$, so that
the critical value $\betacr\approx0.008$. In
Figure~\ref{fig:crit_beta}, we show snapshots of the velocity
structure for flows with $\betastar=0.0003$, 0.006 and 0.009. For the
two lowest $\betastar$ values (shown in the left and middle panels), we
find that the velocity structure approaches the steady-state transonic
wind, passing smoothly through a sonic point on a time-scale of order
several sound crossing times. For the simulation with
$\betastar>\betacr$, however, we find that the flow never attains a
steady-state form. Instead, the flow remains highly variable, with
both alternating epochs of outflow and inflow occurring over
time-scales of order tens of flow time-scales. After an initial
transient phase, this flow structure approaches a quasi-repetitious
pattern, but no steady-state flow is found on a time-scale of 1000s of
flow time-scales. This result demonstrates that when the flow cannot
pass smoothly through a sonic point, no steady state solutions are
allowed, and the flow must become highly variable in time (as
expected).

\subsection{Properties of the variability in non-driven flows}

\begin{figure}
\centering
\includegraphics[width=\columnwidth]{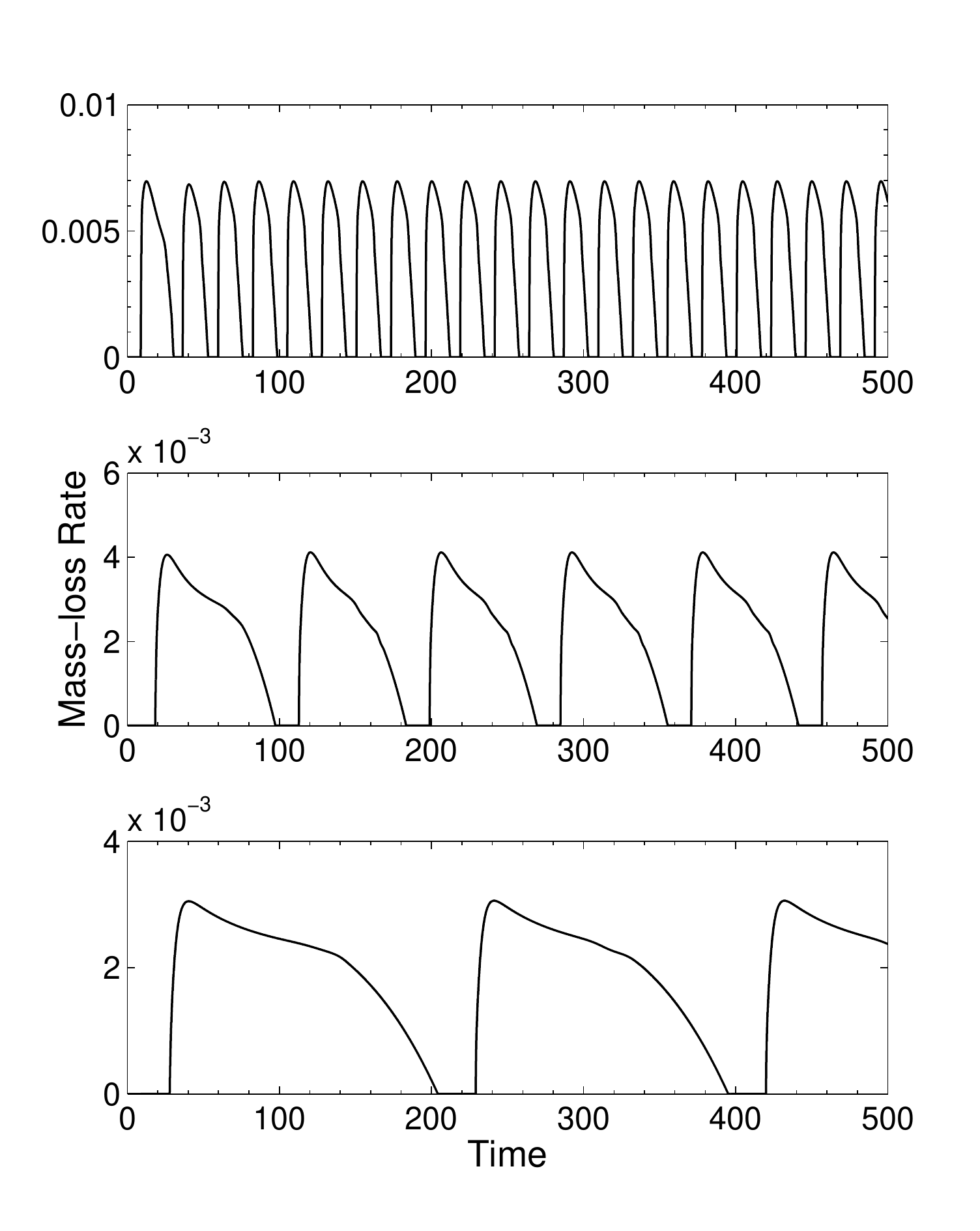}
\caption{Surface mass-loss rate from the planet as a function of time  
for a system with $b=10$, $\theta_s=0.1$, and $\betastar=0.03$. Panels
show simulations with outer boundaries at $\rout$ = 10, 20 and 30
(from top to bottom). Note that the variability time scale depends on 
the value of $\rout$. } 
\label{fig:mdot_boundary} 
\end{figure}

The variability in the flows with $\betastar>\betacr$ settles into a
quasi-repetitious flow with outflow followed by periods of inflow. We
find that the time-scale associated with variability depends on the
size of the domain simulated (i.e., on the radius of the outer
boundary). In Figure~\ref{fig:mdot_boundary}, we show the surface
mass-loss rate $\dot{\Sigma}$ from the planet ($\rho u$ evaluated at
the planet's surface) as a function of time for outer boundaries
$\rout$ = 10, 20 and 30 (from top to bottom). These simulations
use our ``standard'' set of model parameters ($b=10$, $\theta_s=0.1$),
where we use a super-critical value of the stellar background field 
$\betastar=0.03>\betacr$.

\begin{figure}
\centering
\includegraphics[width=\columnwidth]{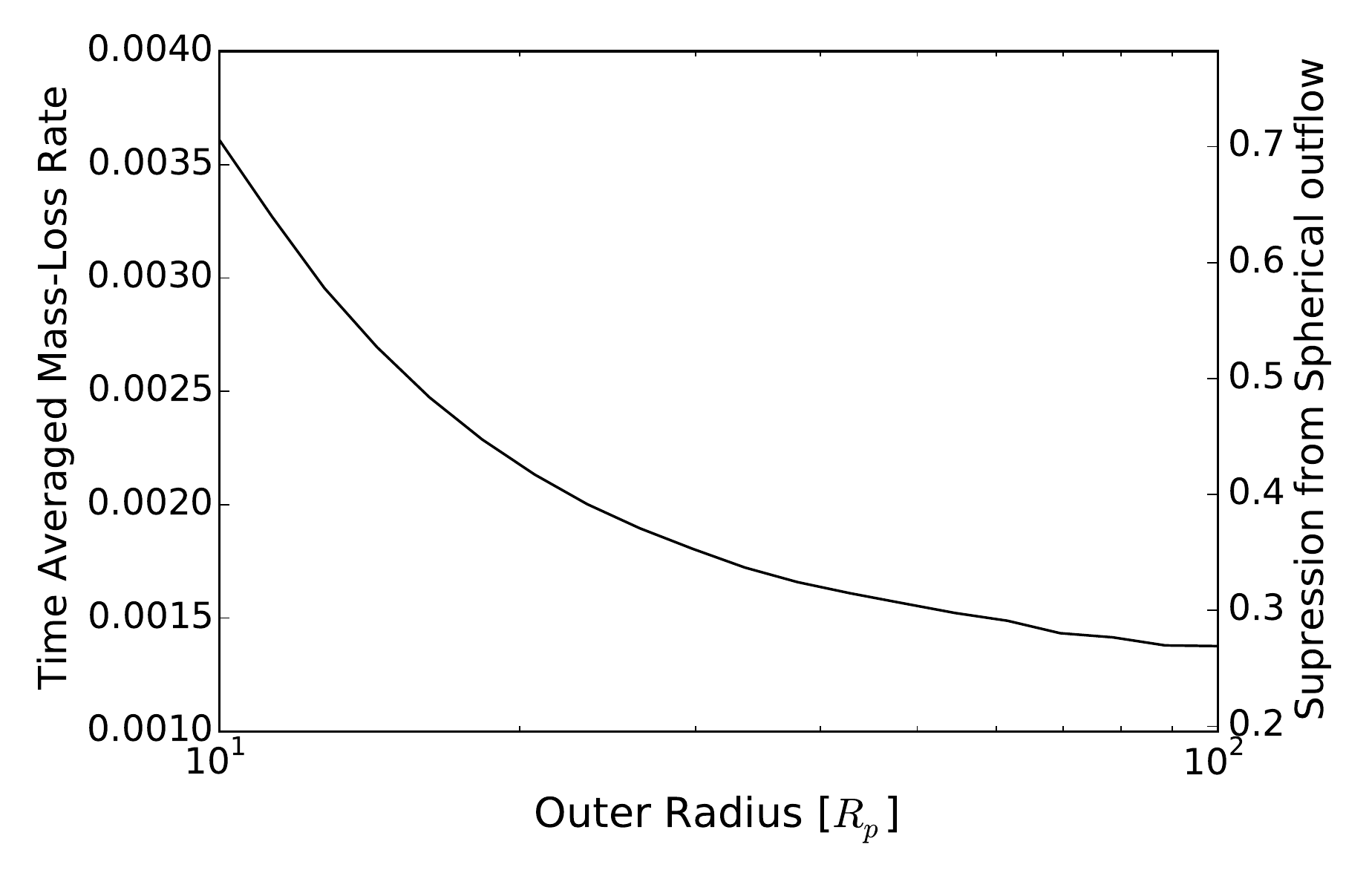}
\caption{The time averaged mass-loss rate (in dimensionless units) as 
a function of outer boundary radius $\rout$ [given here in units  
of $\rplan$]. The right-hand $y$-axis shows the amount by which the 
time-averaged mass-loss rate is suppressed compared to a spherical 
transonic outflow. } 
\label{fig:undriven_mdot}
\end{figure}  

\begin{figure}
\centering
\includegraphics[width=\columnwidth]{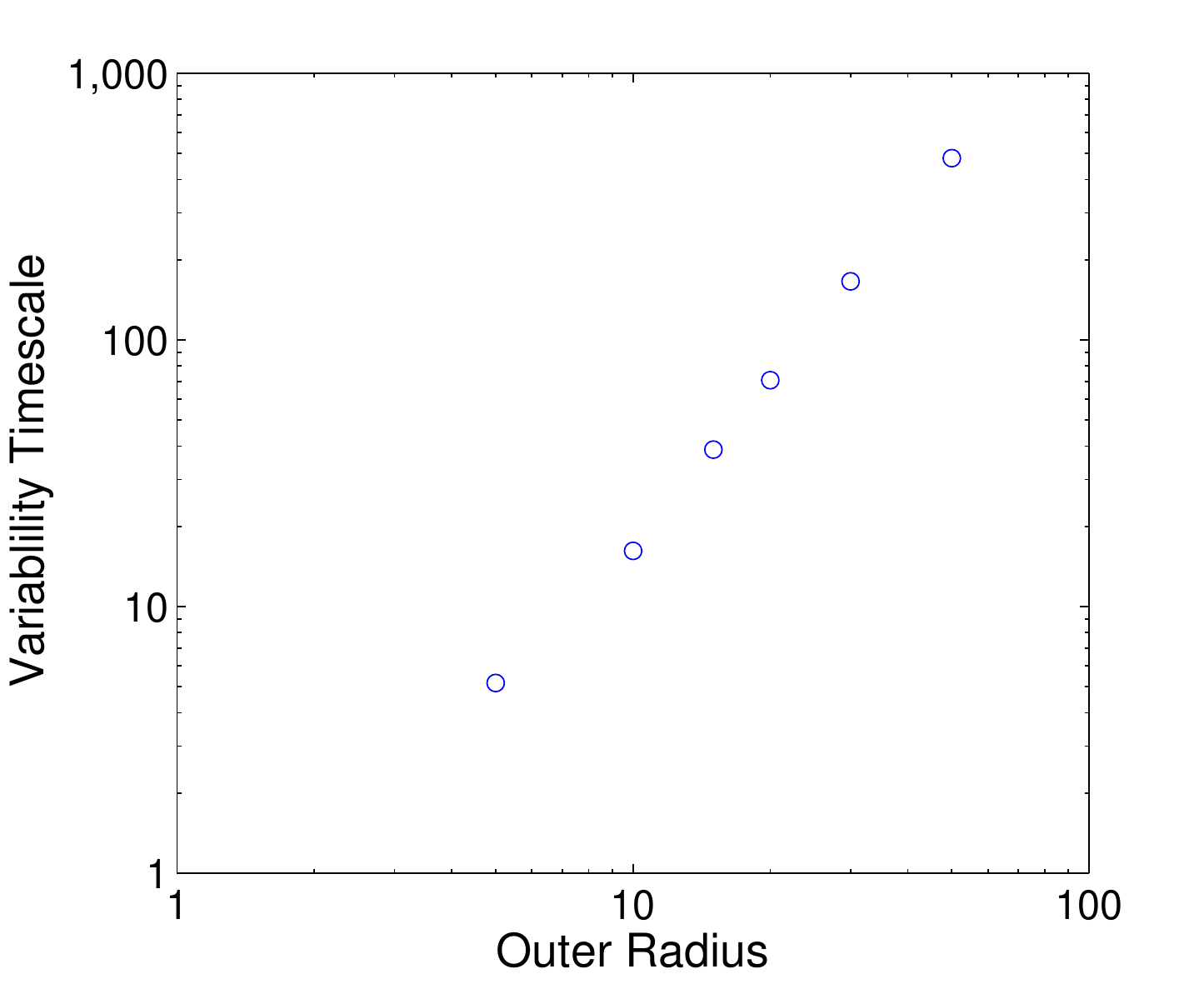}
\caption{Variability time-scale as a function of the outer boundary
radius. The variability time is define as the time interval over 
which $\dot{\Sigma}>0$ at the planetary surface (see Figure 
\ref{fig:mdot_boundary}). Note that the variability time is 
proportional to the the square of the boundary radius. } 
\label{fig:timescale_vs_boundary} 
\end{figure} 

Figure \ref{fig:undriven_mdot} shows the time-averaged mass-loss rate
as a function of outer boundary radius. This plot shows that the
time-averaged mass-loss rate falls steadily with increasing radius of
the computational domain. On the right vertical axis, the figure also
shows the mass-loss rate as a function of its value for a purely
spherical outflow. The mass-loss rate is thus suppressed by a
substantial factor, which varies from about 70\% to only about
25 percent as the outer boundary varies from $\rout$ = 10 to 100.

Figure \ref{fig:timescale_vs_boundary} shows that the variability time-scale
(essentially the period) strongly increases with the radius of the
outer boundary. The simulations also show that the shape of the
surface mass-loss rate profiles appears to be scale free and that the
amplitude of the surface mass-loss profiles decreases with increasing
domain size. Here we define the variability time-scale to be the time
span over which the surface mass loss rate is non-zero, i.e.,
$\dot{\Sigma}>0$. We then plot this time scale as a function of the
outer boundary radius as shown in Figure~\ref{fig:timescale_vs_boundary}.  
This figure shows that the variability time-scale $\tevolve$
increases as a power-law function of the outer boundary radius, where
the approximate scaling law has the form $\tevolve \propto$ 
$\rout^2$. We note this variability time-scale is longer than
the flow time-scale of the entire domain, where this latter time is 
typically a few $\rout$ and obviously increases linearly with 
the outer radius.

\subsection{Model for non-driven flows}

\begin{figure}
\centering
\includegraphics[width=\columnwidth]{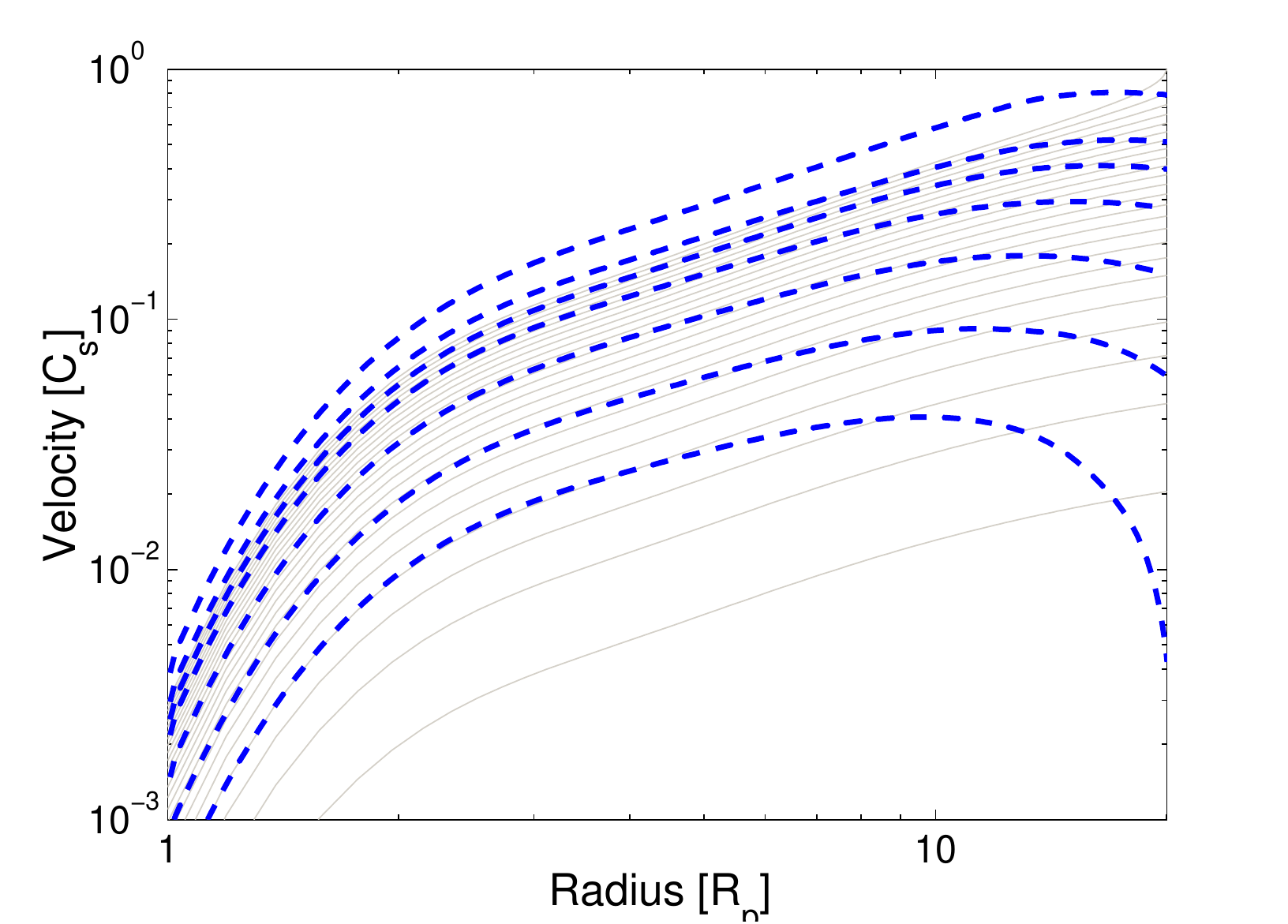}
\caption{Velocity profiles for `breeze' solutions (shown here as the 
thin grey lines) for systems with $\betastar=0.03$, $\theta_s=0.1$, 
and $b=10$, with the outer boundary at 20 $\rplan$. The thick dashed
lines (in blue) show snapshots of the velocity profile from a
simulation with the same parameters. The velocity profile decays in
time, where these snapshots are shown for times of 21.3, 29.0, 36.6,
44.3, 49.6 and 52.7 after the first (uppermost) profile.}
\label{fig:breeze_overlay}
\end{figure}

When the flow is not driven at the inner boundary it develops a
steadily repeating pattern where the flow approaches Mach numbers
$\sim 1$ at the outer boundary but then decays to towards the
hydrostatic solution. It passes through the hydrostatic solution and
inflows until the material at large radius reaches low pressures and
the outflow is launched again. We show snapshots of the velocity
structure at various time-intervals in Figure~\ref{fig:breeze_overlay}
as thick dashed lines.

The variability time-scale is clearly longer than the flow sound
crossing times for the domain $\rout>10$. Thus one might expect
it is possible to setup and maintain a steady flow, over a large part
of the domain. It this regime there are no transonic solutions and the
only available solutions are the breezes - which we note do not
satisfy the Boundary Condition of zero pressure at large radius. These
breeze solutions are plotted in Figure~\ref{fig:breeze_overlay} as
thin grey lines (each line corresponds to a different mass-flux), with
the simulation profiles overlain for $\betastar=0.03$, $b=10$, 
$\rout=20$ and $\theta_s=0.1$.

The profiles from the simulations indicate that the flow is able to
maintain nearly steady-state breeze solutions out to a region close to
the outer boundary. At this boundary, however, the velocity (and hence
the mass-flux) is reduced. If the the mass-flux changes near the
boundary, then clearly the density near the outer boundary must also
be evolving in time. Suppose that we assume that a (nearly)
steady-state breeze solution exists in the inner region close to the
planet; then the density at large radius, near the outer boundary,
will act as the boundary condition that `chooses' which of the breeze
solutions is operating at a given time. 

\begin{figure}
\centering
\includegraphics[width=\columnwidth]{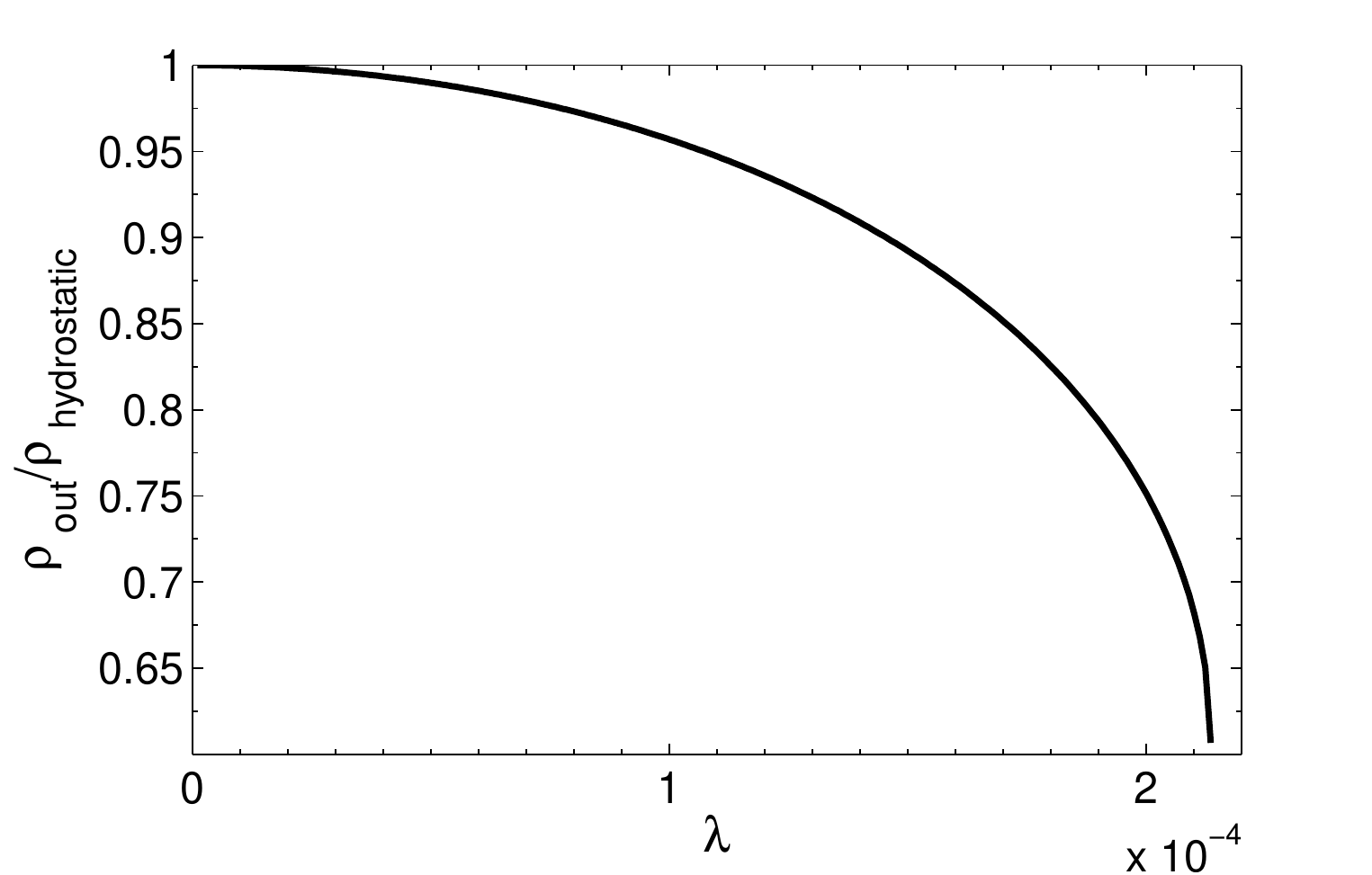}
\caption{Density at the outer boundary as a function of dimensionless  
mass-loss rate $\lambda$, from the solution to Equation 
(\ref{eqn:rho_lambda}), for $\betastar=0.03$, $\theta_s=0.1$ and
$b=10$ with $\rout=20$. No solution exists for sufficiently large
mass-loss rate $\lambda > 2.1351\times10^{-4}$, which corresponds to
$u=\cs$ at the outer boundary. } 
\label{fig:rho_out_lambda}.
\end{figure}

However, because the mass-flux is not constant, but rather decreases
near the outer boundary, the density near the boundary must increase.
Because the breeze solution `chosen' depends on the density at large
radius, the choice of breeze solution will change. As the mass-flux at
the outer boundary is reduced and the density increases, the breeze
solution evolves to cases with lower mass-fluxes. This behaviour occurs
because a higher density at large radius results in a lower
mass-flux. This trend can be seen starting from the steady-state
equations for mass-flux and the Bernoulli potential, i.e.,  
\begin{eqnarray}
\rho u h_\phi h_q &=& \lambda\\ \label{eqn:lambda}
\frac{1}{2}u^2+\log\rho+\psi&=&\epsilon \label{eqn:bernoulli}
\end{eqnarray}
where $\lambda$ is the mass flux and $\epsilon$ is the value of the
Bernoulli potential (and where both $\lambda$ and $\epsilon$ are 
constants). Following \citet{Adams2011} we can define an ancillary 
function $H$ such that $h_q h_\phi = q/H^{1/2}$ and 
\begin{equation}
H=(\betastar+2/\xi^3)^2\cos^2\theta + 
(\betastar-1/\xi^3)^2\sin^2\theta\,. 
\end{equation} 
Applying our dimensionless boundary conditions that $\rho=1$ and
$u=u_1$ at the planet's surface with radius $\xi=1$, and taking
$\rho=\rho_{\rm out}$ at the outer boundary $\xi=\rout$,
Equations~(\ref{eqn:bernoulli}) can be solved to derive an 
expression for $\rho_{\rm out}$ in terms of the dimensionless 
mass flux $\lambda$,  
\begin{equation}
\rho_{\rm out}=\frac{\lambda H_{\rm out}^{1/2}}{q}\frac{1}{\cal D}\,,
\label{eqn:rho_lambda}
\end{equation}
where we have defined 
$${\cal D} \equiv 
\sqrt{-W_0\left[-\lambda^2H_{\rm out}/q^2\exp
\left(-b/\rout+b-\lambda^2H_1/2q^2\right)\right]}\,,$$
where $W_0$ is the Lambert $W$-function. The solution to Equation
(\ref{eqn:rho_lambda}) is plotted in Figure~\ref{fig:rho_out_lambda},
where $\rho_{\rm out}$ is scaled to the hydrostatic value, i.e., the
value corresponding to no outflow and $\lambda=0$; the other system
parameters are taken to be $\betastar=0.03$, $\theta_s=0.1$, and
$b=10$, with the outer boundary located at $\rout=20$.

As shown in Figure \ref{fig:rho_out_lambda}, a larger density
$\rho_{\rm out}$ at the outer boundary results in a lower mass-flux
$\lambda$. Given this dependence, we argue that breeze solutions with
one-way outer boundaries (such that material can flow outward, but not
back inwards) are subject to an instability.  Consider a steady state
`breeze' solution that causally connects the outer boundary and
planet's surface. If one perturbs the mass-flux at large radius on a
time-scale longer than the flow time-scale, this action will result in
a density change at large radius due to mass continuity. Since
$\rho_{\rm out}$ is a monotonically decreasing function of $\lambda$,
this process will be unstable: Suppose that the perturbation increases
the density at large radius; the resulting mass flux will tend to
follow the curve in Figure \ref{fig:rho_out_lambda} and hence
decrease, which will increase the density even further. We suspect
that this instability is similar to the one described by
\citet{Velli1994,DelZanna1998} for sub-sonic spherical outflows,
resulting in a hysteresis-like cycle of outflow and inflow.

Taking this instability to be the cause of the cycle seen in our
simulations, one can build a simple model that aims to reproduce the
time-scale of variation and the self-similar mass-flux profile, along
with its amplitude as seen in Figure~\ref{fig:timescale_vs_boundary}.
Toward this end, we can write the mass contained in the flow at large
radius in the form: 
\begin{equation}
M_{\rm out}\propto\int^t \lambda\, {\rm d}t \,.
\label{flowmass} 
\end{equation}
If we now assume that the decrease in mass-flux is sensitive to the
outer boundary, then the region over which the flow `feels' the outer
boundary should be comparable to the scale height in the flow (denoted
here as $\hscale$).  Equation (\ref{flowmass}) can be rewritten as: 
\begin{equation}
\rho_{\rm out}h_q(\rout) h_\phi(\rout) \hscale(\rout) 
\propto \int^t \lambda\, {\rm d}t\,.
\end{equation}
We can turn the above expression into a differential equation for the
evolution of $\rho_{\rm out}$ simply by taking the time derivative, 
so that it becomes:  
\begin{equation}
\frac{d \rho_{\rm out}}{d t}h_q h_\phi \hscale \propto \lambda \,. 
\label{eqn:diff_eqn}
\end{equation}
Since $\lambda$ depends on the density at the outer boundary through
Equation~(\ref{eqn:rho_lambda}), one can find a single differential
equation to solve for the evolution of $\rho_{\rm out}$.

Equation~(\ref{eqn:diff_eqn}) can be solved numerically. However, for
non-zero values of $\betastar$, and for large boundary radii where
$\rout\gg 1$, Equation~(\ref{eqn:rho_lambda}) is actually
roughly independent of $\rout$. Note that the $b/\rout$
term is small compared to $b$, and that at large radius $H_{\rm out}$
approaches a constant value: 
\begin{equation}
H \rightarrow\betastar^2 \quad \mbox{ as } \quad \xi\rightarrow\infty\,.
\end{equation}   
This limiting form arises because the magnetic field lines are the
nearly vertical with zero divergence at large radius (as they are
dominated by the stellar background contribution). This result is not
surprising, in that the domination of the stellar background field is
the origin of the complication that that transonic flows do not exist
in the first place.

In the limiting regime described above, the density $\rho_{\rm out}$
is almost purely a function of the dimensionless mass flux $\lambda$. 
The differential equation itself can easily be made dimensionless,
allowing us to extract a time-scale scaling with $\rout$. Noting
$qH^{-1/2} =h_qh_\phi$, we find that the $h_q h_\phi$ term on the
left-hand-side of the differential equation cancels. The time-scale
for the evolution of the differential equation is then given by: 
\begin{equation}
\tevolve \sim \hscale \,.
\end{equation} 
Since the breezes are sub-sonic, the density structure of the flow can 
be considered as nearly hydrostatic, which implies that the scale 
height is given approximately by:  
\begin{equation}
\hscale(\rout)=\left|\frac{\partial \log \rho}{\partial r}
\right|^{-1}=| g|^{-1} = \rout^2/b \,.
\label{scaleheight} 
\end{equation}  
Thus, the $R^2$ dependence of the time-scale comes from the fact the
scale height at large radius scales as $R^2$ and the background
stellar field forces the stream-bundle area to be constant. Notice
also that --- in addition to the scaling --- the absolute value for
the variation time-scale is comparable to that found in
Figure~\ref{fig:timescale_vs_boundary} for $b=10$. For completeness, 
we note that the scale height in equation (\ref{scaleheight}) does 
not take into account the geometry of the magnetic fields. In the limit 
$\xi \gg 1$, the field lines are nearly vertical, and the scale height 
should be corrected by the geometrical factor $\xi/z$. In the extreme 
limit, however, $\xi/z\to1$, so we can ignore this complication. 

Curiously, because the scale height increases as $\rout^2$, for
sufficiently large domain sizes $\rout$ the evolution time is longer
than the sound crossing time of the flow, where $t_{\rm cross}=\rout$
in dimensionless units.  This finding suggests that only a fraction of
the flow will be able to adjust during a crossing time, where this
fraction is given by $f=t_{\rm cross}/t_{\rm evolve}=b/\rout$.  For
the case of our example shown in Figure~\ref{fig:breeze_overlay}, this
fraction $f = 1/2$. Note that the flow profiles from the simulation
match onto the breeze solutions for the inner half of the domain and
diverge from the breeze solutions in the outer half.
 
\begin{figure*}
\centering
\includegraphics[width=\textwidth]{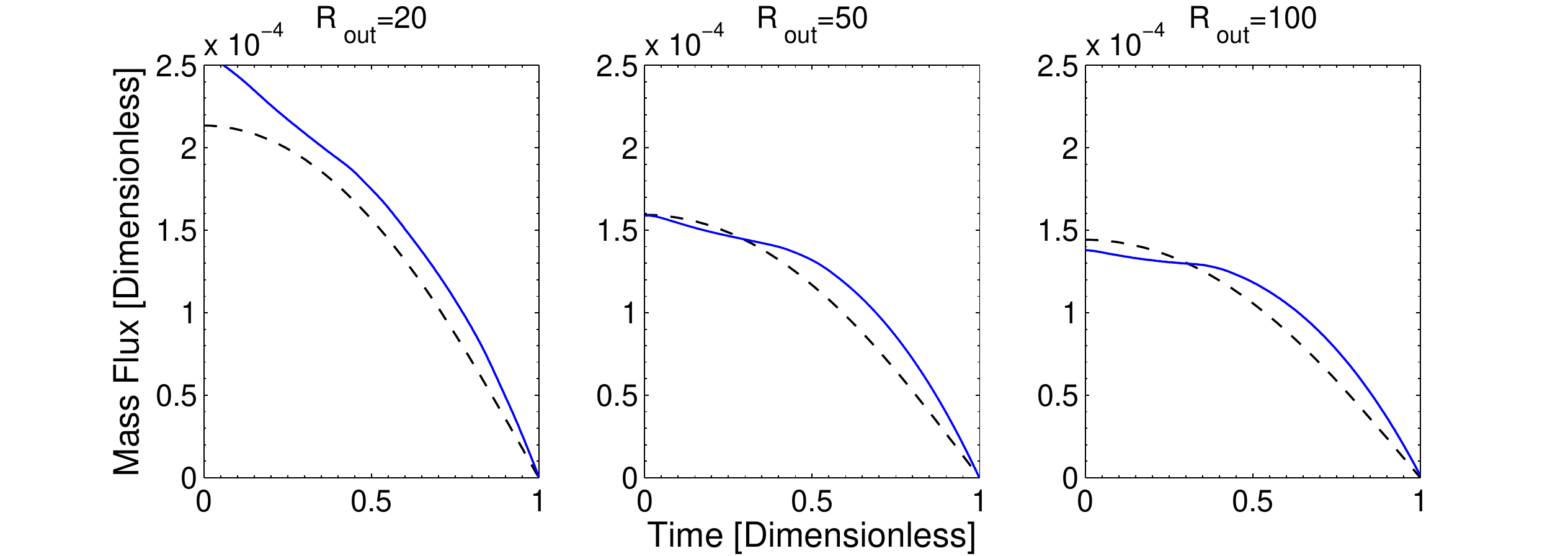}
\caption{Evolution of mass-flux versus time for numerical calculations
(solid curve) and a simple model described in the text (dashed curve). 
The time variable (horizontal axis) is scaled so that the total time 
evolution is unity in all cases.} 
\label{fig:lambda_time} 
\end{figure*} 

A final test is to predict the evolution of time-varying outflow
solution as it evolves by sampling through various breeze solutions;
we would also like to understand the amplitude of these variations. In
the limit $\rout\gg 1$, the solution must become self-similar,
as the $b/\rout$ term can be dropped and quantity $H_{\rm out}$
approaches a constant. Assuming that the solution starts with speed
$u_{\rm out}=1$, we can numerically integrate the differential
equation to find the time evolution of the mass-flux ($\lambda$). The
solutions are plotted in Figure \ref{fig:lambda_time} for several
values of $\rout$; these results are compared to the numerical
solutions, for which the mass flux $\lambda$ is measured at 
$\rout/2$.  The agreement is good, in both amplitude and
evolution. We can also check the time-evolution for different values
of $q$ and $\betastar$; the solutions should independent of these
values, provided the systems are in the regime which does not allow
transonic flow. Which indeed we find with $\betastar$ range 0.02 to
0.2 and $\theta_s$ in range 0.01 -- 0.2.

The discussion above indicates that the origin of the time-variability
in undriven magnetically controlled flow is an instability in the
breeze solutions; moreover, the time-scale of the variations is set by
the scale height at the outer boundary such that $\tevolve$
$\propto \rout^2$. Note that since the velocity is not zero at
the outer boundary, one must write Equation (\ref{eqn:diff_eqn}) as a
scaling relation rather than an equality. All of the subsequent
analysis assumes that the relevant constant is roughly independent of
the parameters of the simulation. The simulations seem to suggest that
this is true. We hypothesis that are full linear analysis of the 
{\it global} modes of any instability might give further insight.
Since this type of work has not been performed for even simple
spherical outflows due to the lack of obvious basis functions 
(for a discussion of possible approaches to spherical breezes, see
\citealt{Theuns1992}), we shall not attempt such an analysis here.

\section{Driven Flow}
\label{sec:driven}

Time variations in the outflow can arise for two conceptually
different reasons.  In the first case, as outlined in the previous
section, time dependent flow arises when the fluid cannot pass
smoothly through a sonic transition (and this type of time variation
occurs even for steady heating). In addition, planets can also
experience external variability, so that outflows can be time
dependent due to external factors. In this case, the time scale of the
variability will be linked to the driving mechanism.  Possible sources
of such driving variations include stellar flares or outbursts, and/or
an eccentric orbit that modulates the flux of UV and X-ray radiation
received by the planet. Any modulation in the flux will change the
density at the base of the outflow. For simplicity, we consider two
types of driving: Sinusoidal driving and pulsed driving, where we vary
the density at the base of the flow (in the ghost zones) as a function
of time.

\subsection{Sinusoidal Driving}

To illustrate the effects of driven, time-dependent forcing, we
consider the following simple model where the density at the base of
the flow varies as: 
\begin{equation}
\rhobase=A\sin^2\left(\frac{t}{T_{\rm var}}\right)+1\,,
\label{sinedrive}
\end{equation}
where $A$ is the amplitude of the variation and $T_{\rm var}$ is the
time-scale on which it occurs. 

\begin{figure*}
\centering
\includegraphics[width=0.8\textwidth]{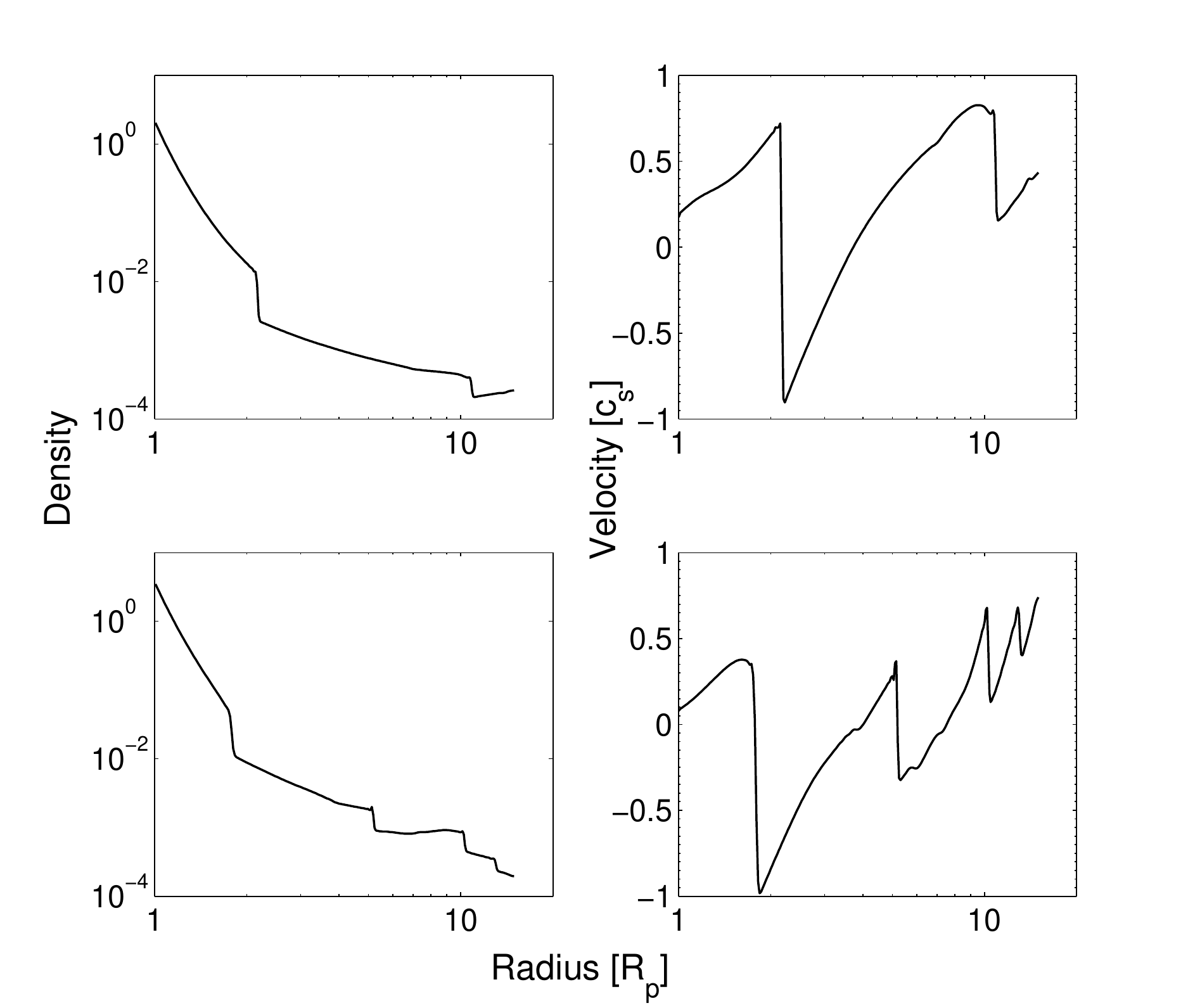}
\caption{The density (left panels) and velocity (right panels)
structure of the flow for a variability time-scale of $T_{\rm var}=2$
(top) and $T_{\rm var}=1$ (bottom). The simulation parameters are
$b=10$, $\betastar=0.01$, $\theta_0=0.1$, and the domain has an outer
radius $\rout$ = 15.}
\label{fig:sine_snapshot}
\end{figure*}

We have performed a collection of numerical simulations using Equation
(\ref{sinedrive}) as an inner boundary condition. After some initial
transients, the flow settles into a quasi-repetitious pattern.
Figure~\ref{fig:sine_snapshot} shows snapshots of the density and
velocity structure of outflows with variation time scale 
$T_{\rm var}=2$ (top panels) and $T_{\rm var}=1$ (bottom panels). The~
flow structures depicted in Figure~\ref{fig:sine_snapshot} show some
generic features that are present in all of the simulations. Both the
density and the velocity distributions show a series of
discontinuities that originate as weak shocks near the planet and then
propagate outwards.

\begin{figure}
\centering
\includegraphics[width=\columnwidth]{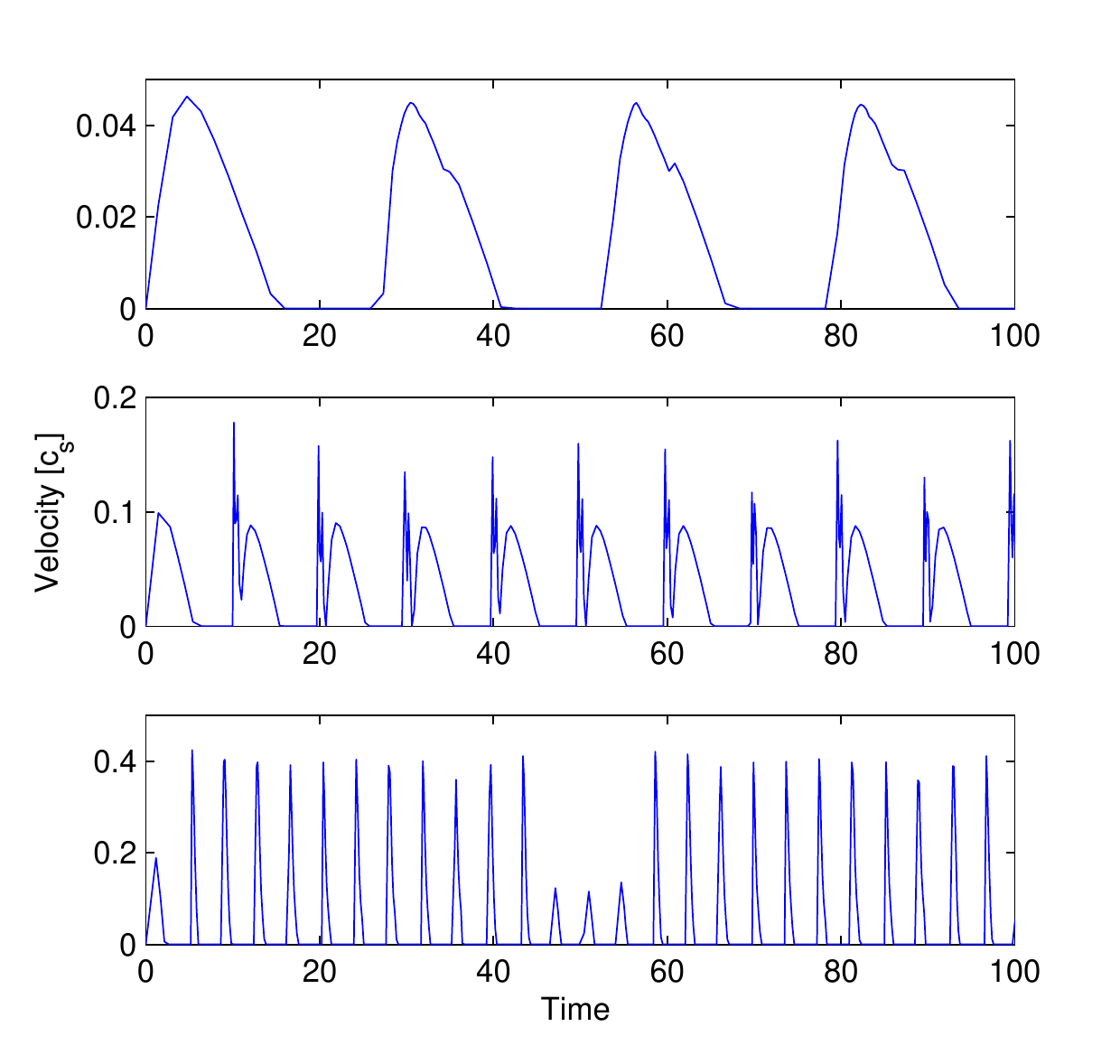}
\caption{Velocity at the planet surface as a function of time for  
time-varying, driven flow with amplitdue $A=3$ and time scale 
$T_{\rm var}$ = 8.25 (top), 3.83 (center), and 1.78 (bottom). The
system parameters are taken to be $\betastar=0.03$, $\theta_s=0.1$,
and $b=10$. } 
\label{fig:var_drive}
\end{figure}

We now turn our attention to the outflow properties as a function of
the driving time-scale $T_{\rm var}$.  In Figure~\ref{fig:var_drive},
we show how the velocity at the planet various with time for three
different driving time-scales of $T_{rm var}$ = 8.25, 3.83, 1.78. The
figures shows that the amplitude of the velocity variations decreases
with increasing variation time-scale (note the different scales on the
vertical axis in the three panels). Furthermore, while the time-scale
of the velocity variations is direct mapping of the driving
time-scale, the profiles are not symmetric. For the fastest driving
time-scale (1.78, bottom panel), after a time of roughly 50 the
amplitude of the variation drops by a factor of $\sim 3$ and this
variations of a time-scale of 50 exists for the entire simulation
(over a time of several 1000s).

\begin{figure}
\centering
\includegraphics[width=\columnwidth]{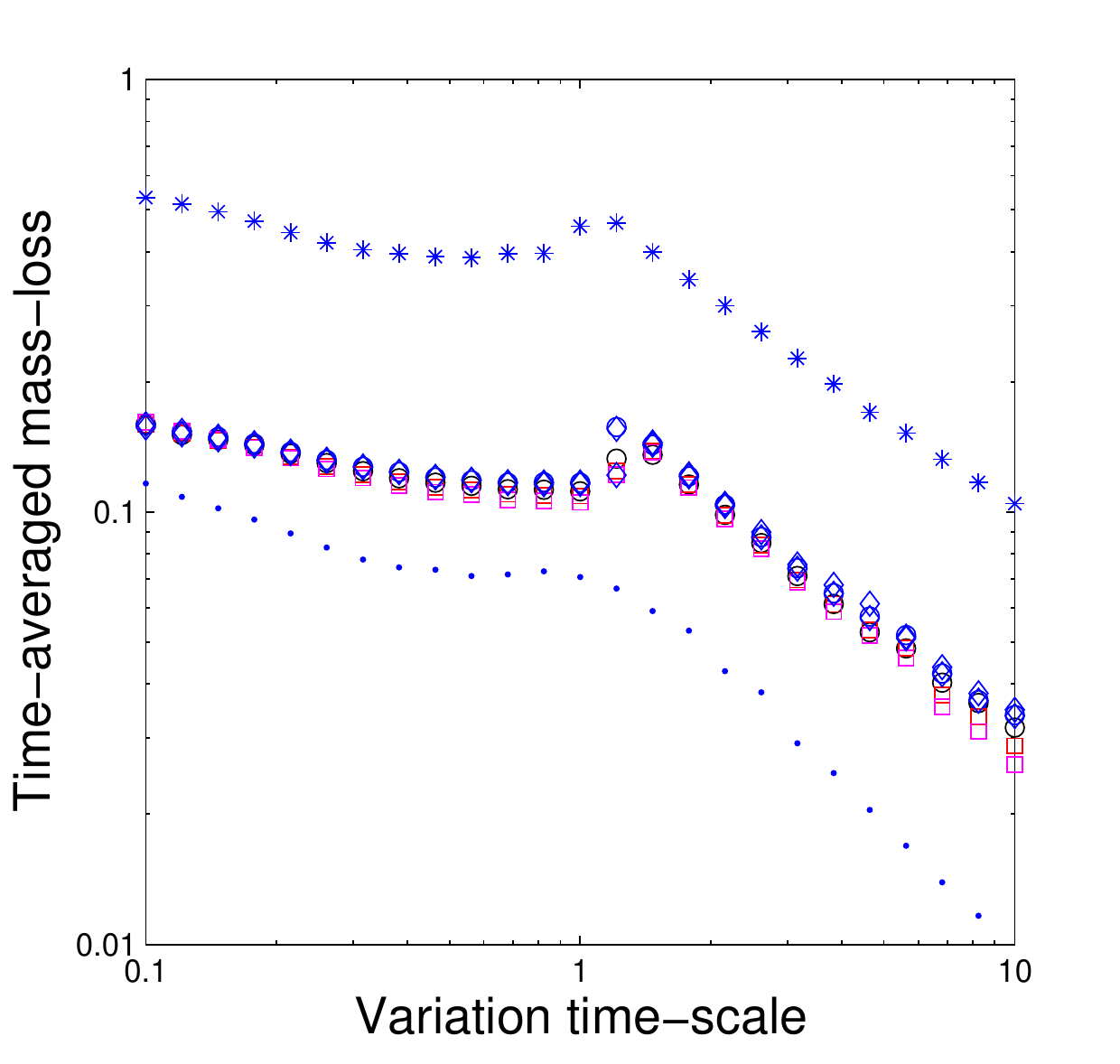}
\caption{The time-averaged mass-loss rate as a function of driving 
time-scale. The open points show results for $b=10$ and $A=3$, where
the other parameters are allowed to vary. The stars show our
``standard'' model but with $A=10$ (upper curve), whereas the dots
show our standard model but with $b=15$ (lower curve). } 
\label{fig:avg_mdot}
\end{figure}

In Figure~\ref{fig:avg_mdot} we show the time-averaged surface
mass-loss profile for a range of driving time-scales in the range
$T_{\rm var} = 0.1-10$, where the averaging has taken place over a
simulation time of 500.  As shown in the Figure, the time-averaged
mass loss rate is relatively constant as a function of the driving
time for the regime where $T_{\rm var} < 1$, and decreases for larger
values of $T_{\rm var}$.  More specifically, the mass loss rate
displays a moderate peak near $T_{\rm var}\approx1$, and decreases
according to $1/T_{\rm var}$ for large driving time scales. 

We can understand the behaviour shown in Figure \ref{fig:avg_mdot} as
follows.  For small values of the driving time, smaller than the sound
crossing time, the flow cannot adjust to the changing driving density.
As a result, in this regime the driving density (see equation
[\ref{sinedrive}]) can be effectively replaced with a time-averaged
value. However, the numerical simulations also show that the flow is
not able to cycle through its low-mass-loss states, as it does in the
case of truly constant base density (see Figure \ref{fig:mdot_boundary}). 
As a result, the mass loss rate is more nearly constant in time
(instead of showing periodic behaviour) and the time-averaged value is
correspondingly higher.  In the opposite regime, where the driving
time scale is much longer than the crossing time, the flow has time to
adjust and reach its low states, so that the mass loss rate decreases.
In addition, the driving density (see equation [\ref{sinedrive}]) is
only larger than its baseline value (chosen here to be unity) for a
fraction of the time, where the fraction $\propto 1/T_{\rm var}$.

In Figure \ref{fig:avg_mdot}, the points show simulation results with
$b=10$ and $A=3$ held fixed and varying values of $\betastar$ and
$\theta_s$. The stars stars show our ``standard'' case but with an
increased amplitude of $A=10$; the dots show simulations with $b=15$.
We note the profiles are somewhat similar and have a power-law
fall-off once the variation-time-scale is greater than unity, where
the time-averaged mass-loss rate can approximately be described as
$\propto T_{\rm var}^{-1}$.

\subsection{Pulsed Driving}

\begin{figure*}
\centering
\includegraphics[width=\textwidth]{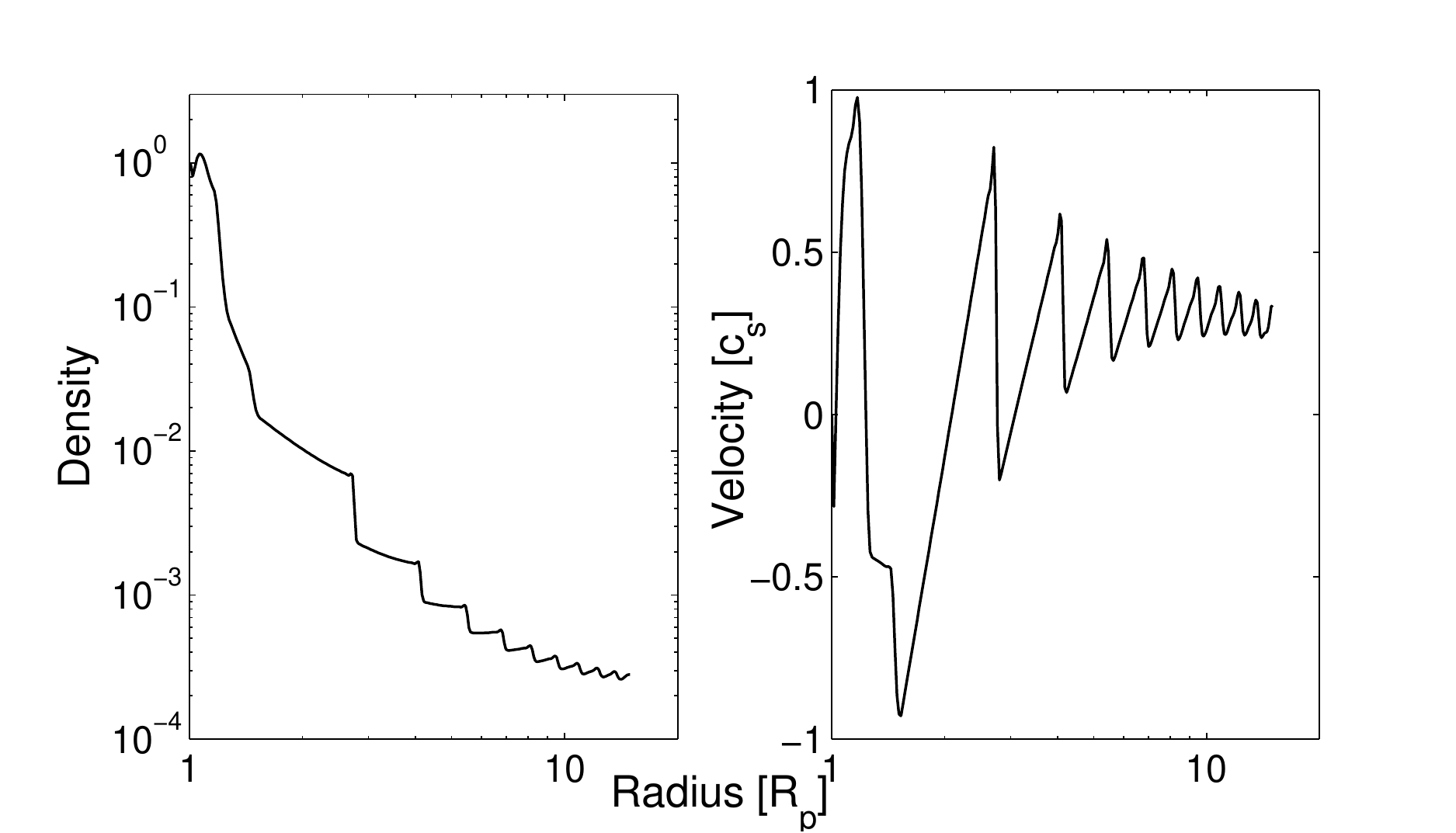}
\caption{The density (left panel) and velocity (right panel)
structure of the outflow for pulsed driving. The variability time-scale 
$t_{\rm var}=1$ and the pulse duration $\Delta_T=0.1$. The simulation 
parameters are $b=10$, $\betastar=0.01$, and $\theta_0=0.1$, with an 
outer radius of the domain at $\rout$ = 15. } 
\label{fig:pulse_snap}
\end{figure*}

Another possible source of driven time variations is UV and X-ray
heating from stellar flares. These bursts of radiation could act to
heat the upper layers of the planetary atmosphere and thereby increase
the density at the base of the outflow. Flares act in a pulsed manner,
which can be characterized by a time scale $\Delta_T$ over which the
radiation is enhanced and by an amplitude $A_p$ that specifies the
degree of enhancement. In this set of simulations, we also let the
pulses repeat on a time-scale of $t_{\rm var}$. In
Figure~\ref{fig:pulse_snap}, we show a snapshot of the flow properties
for our ``standard'' set of parameters with 
$\Delta_T=0.1$ and $t_{\rm var}=1$. 

We note that this figure shows significant similarities with the flow
variations arising from the sinusoidal driving case discussed in the
previous subsection. Specifically, discontinuities in the flow develop
close to the planet due to weak shocks and then propagate outwards. As
a result, the qualitative features we can expect in the case of driven
flow are not particularly sensitive to the exact form of the driving,
but depend primarily on the driving time-scale.
 
\begin{figure}
\centering
\includegraphics[width=\columnwidth]{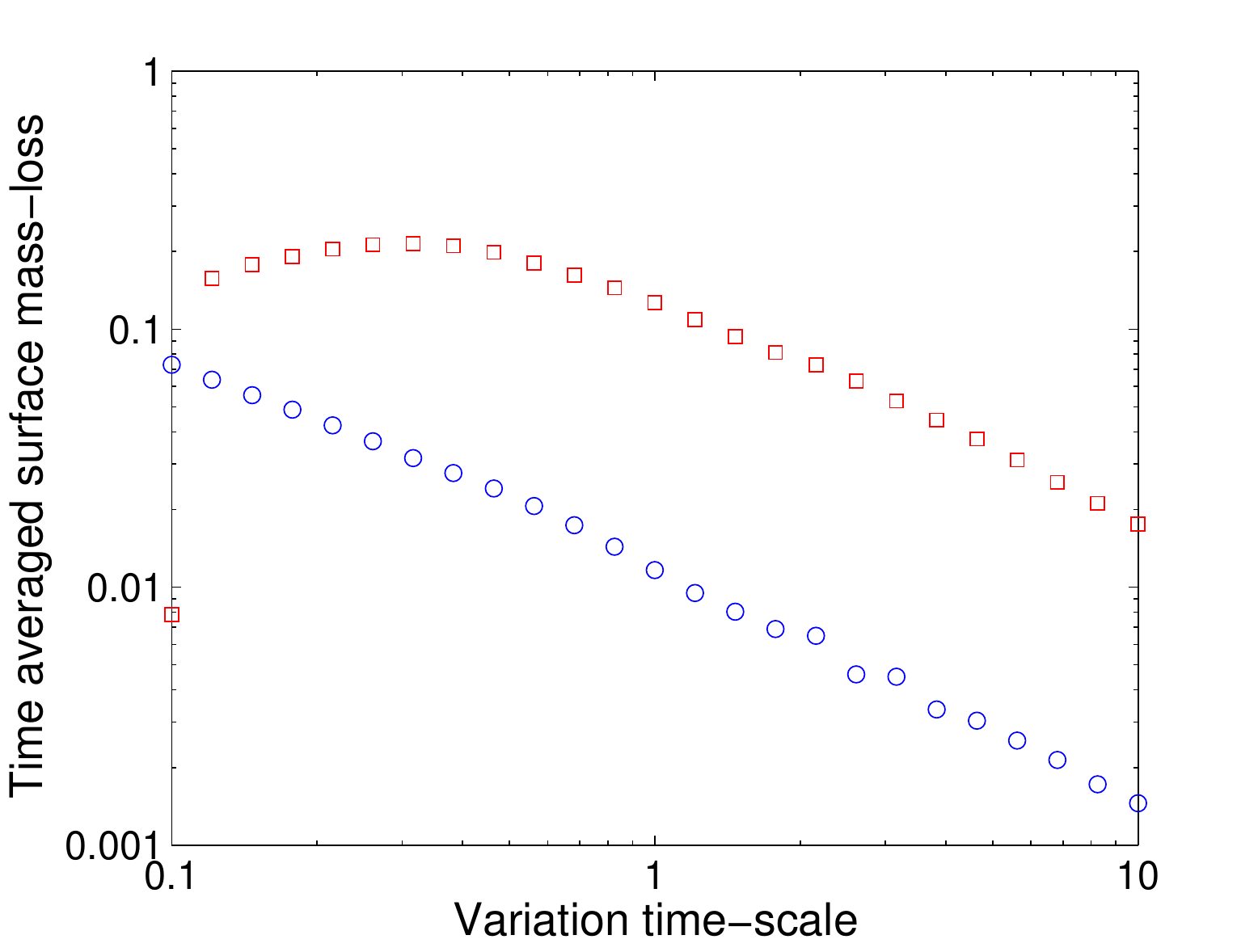}
\caption{Time averaged mass-loss rates for the case of pulsed driving  
terms for pulse durations of $\Delta_T=0.1$ (red squares) and
$\Delta_T=0.01$ (blue circles). Mass-loss rates are shown as a function 
of the time scale $t_{\rm var}$ on which the pulses repeat. } 
\label{fig:avg_mdot_pulse}
\end{figure}

Once again we turn our attention to the general outflow properties as
a function of the driving time-scale, here the time $t_{\rm var}$ on
which the pulses repeat. Figure~\ref{fig:avg_mdot_pulse} shows the
time-averaged surface mass-loss rate as a function of the variation
time-scale $t_{\rm var}$ for systems driven with pulses of duration
$\Delta_T=0.1$ (red squares) and $\Delta_T=0.01$ (blue circles). As
shown in the figure, when the driving time-scale is significantly
longer than $\Delta_T$, we see an approximate power-law fall off in
time-averaged mass-loss rate with variation time-scale. This variation
can be approximated as a power-law of the form $\propto\Delta_T^{-1}$,
analogous to the behavior found for sinusoidally driven flows.

\section{Implications} 
\label{sec:imp} 

\subsection{Flow Regimes for Observed Exoplanets} 

For the case of magnetically controlled flow, the magnetic field
topology can prevent the flow from attaining a steady-state solution, 
in part due to the inability of the flow to pass smoothly through the
sonic transition.  This regime arises for magnetically controlled
evaporation of close-in exoplanets when the background stellar field
is sufficiently strong. For the case of anti-aligned dipoles on the
star and planet, the criterion for unsteady outflow \citep{Adams2011}
is given by the expression 
\begin{equation}
\frac{B_*}{\bplan}\left(\frac{R_*}{a}\right)^3
\left(\frac{G\mplan}{\rplan\cs^2}\right)^3=\betastar{b^3}\gtrsim8\,.
\label{eqn:vari_crit}
\end{equation} 
For a Hot Jupiter orbiting a Sun-like star with surface field 
strength $B_*\approx 1$ G in a 3~day orbit, we find that 
variable magnetically controlled outflow will arise when the 
planetary field falls below the limit: 
\begin{eqnarray}
\bplan&\lesssim& 0.4~{\rm G}~\left(\frac{B_*}{1~\mbox{G}}\right)
\left(\frac{R_*}{1~{\rm R}_\odot}\right)^3
\left(\frac{a}{0.04~{\rm AU}}\right)^{-3}\nonumber\\
&\times&\left(\frac{\mplan}{1~{\rm M_J}}\right)^3
\left(\frac{\rplan}{1.4~{\rm R_J}}\right)^{-3}
\left(\frac{\cs}{10~{\rm km s}^{-1}}\right)^{-6}\,. 
\end{eqnarray}
Simulations carried out previously \citep{Owen2014f} indicate that
magnetically controlled flow will arise for planetary field strengths
$\bplan\gtrsim 0.1$~G. The window for evaporating planets where the
planetary field is simultaneously strong enough for magnetically
controlled flow and weak enough for the stellar background field to
compromise steady-state flow is thus 0.1~G $\lesssim B_P \lesssim$
0.4~G.  Time-variations will thus arise for planets in orbit around
older main-sequence stars when the planetary field is moderately weak, 
or when the planet is extremely close to the star.

On the other hand, for a young Hot Jupiter around a pre-main-sequence
Sun-like star (of age $\sim$ 1-10 Myr), the stellar magnetic fields
are typically much stronger, $B_\ast \sim 1$~kG (e.g.,
\citealt{johnskrull}). The above criterion indicates that the flow
will be time-variable for planetary field strengths $\bplan\lesssim
475$~G. For young star/planet systems, a great deal of parameter space
exists for which the planetary field is strong enough to control the
flow, but weak enough so that the stellar background field dominates
enough to compromise steady-state solutions. As a result, evaporating
planets around young stars are likely to experience variability of the
kind we describe here. 

\begin{figure}
\centering
\includegraphics[width=\columnwidth]{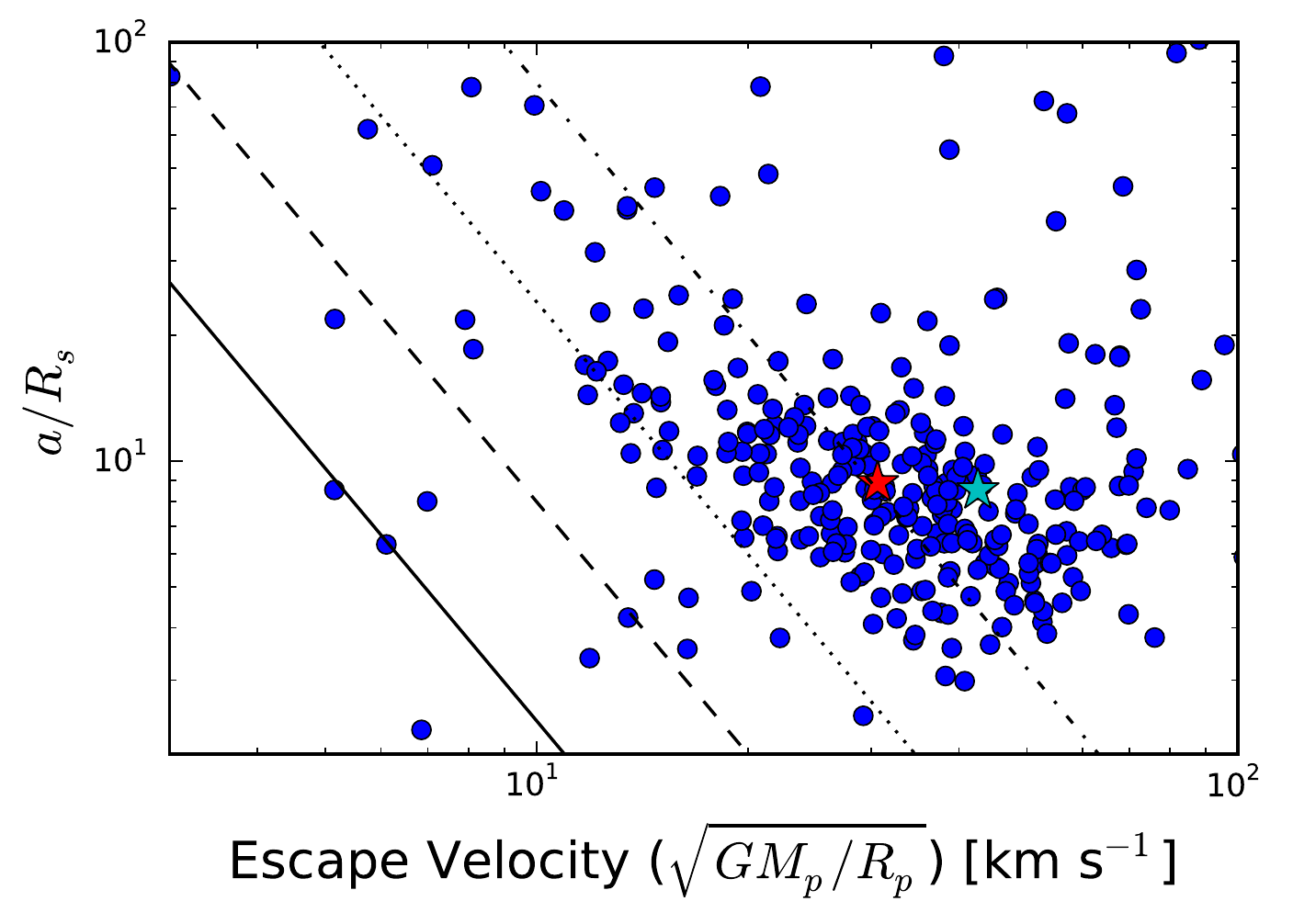}
\caption{The ratio of semi-major axis to stellar radius plotted  
against the escape velocity from the planetary surface for observed
exoplanets (shown as points). The lines show a constant stellar to
planetary surface magnetic field strengths (0.3 -- solid, 1 -- dashed,
3 -- dotted, and 10 -- dot-dashed) assuming sound speed $\cs=10$ km
s$^{-1}$. For a given ratio of field strengths, planets that fall
below the line are susceptible to non-steady-state outflows. The star
symbols represent currently observed evaporating Hot Jupiters HD
209458b (red) and HD 189733b (cyan). }
\label{fig:compare}
\end{figure}

We compare the criterion for magnetically controlled evaporation
variability given in Equation~(\ref{eqn:vari_crit}) to the exoplanet
data in Figure~\ref{fig:compare}. Specifically, we plot the ratio of
the semi-major axis to the stellar radius against the escape velocity
($\sqrt{G\mplan/\rplan}$) for a collection of observed exoplanets
(where the observational values are taken from the Open Exoplanet
Catalogue on 31/08/2015; \citealt{Rein2012}). The figure shows lines
of constant stellar-to-planetary surface magnetic field strength,
where the ratio has the value 0.3 (solid), 1.0 (dashed), 3 (dotted), 
and 10 (dot-dashed).  Planets that lie below the line (for a given
field strength ratio) could be prone to variability. Note that for
early evolutionary times, the planets will have larger radii (as
planets cool and shrink with time), so that the points will move to
the left for younger planet populations.

Figure~\ref{fig:compare} shows that if Hot Jupiters have strong enough
magnetic fields to be magnetically controlled, $\bplan\gtrsim 0.1$~G
\citep{Owen2014f}, then a significant fraction of them are likely to
experience time-dependant evaporation. However, we note that the two
observed evaporating planets HD 209458b and HD 189733b (marked by
stars in Figure~\ref{fig:compare}) are unlikely to be experiencing
variability of this kind. In order for the planets to have weak enough
magnetic fields for a 1~gauss stellar field to produce the required
ratio, the planetary field would be too weak to control the flow;
alternately, the high ratio of stellar to planetary field could be
achieved by strong fields on the stellar surface, but they are rare.

Since stellar magnetic fields are expected to be considerably larger
at early times, up to 1000 times stronger, we suspect that almost all
Hot Jupiters will experience evaporation variability of this kind
provided the flow is magnetically controlled.

\subsection{Physical time-scales}

For convenience, almost all of the work in this paper has been
performed in dimensionless units. In order to discuss observational
implications, however, we must consider typical physical parameters
for the systems of interest. Here we consider separately the two
regimes of variability, those for undriven and driven outflows.

\subsubsection{Undriven flow}

For undriven flow we found that the time-scale for variability in
dimensionless parameters is $\rout^2/b$, which can be written 
in physical variables in the form: 
\begin{equation}
t_{\rm var}=\frac{\rout^2\cs}{G\mplan}\approx 6~\mbox{h~}
\left(\frac{\rout}{20 R_{\rm J}}\right)^2
\left(\frac{\cs}{10~\mbox{km s}^{-1}}\right)
\left(\frac{\mplan}{1~{\rm M}_J}\right)^{-1}\,.
\label{eqn:tphysical}
\end{equation}  
Obtaining the actual time-scale depends on specifying the choice of
the `outer boundary'. In conceptual terms, the outer boundary of the
flow regime is set by the outer boundary of the effective sphere of
influence of the planet.  Given that the flow is magnetically
controlled, one sensible choice would be the radius at which the
stellar magnetic field starts to curve back towards the star and can
no longer be considered approximately vertical. This transition will
occur on a length scale of $(a/R_*)\rplan$, which falls in the range of
$5-100$ times the planetary radius. The corresponding time scales for
such boundary radii vary from a few hours to a few days, variability
times that are eminently observable.

Another choice for the outer boundary scale is the radius at which the
planetary wind becomes entrained in the stellar wind: When the outflow
from the planet is sufficiently far from the surface, the ram pressure
of the stellar wind will dominate that of the planetary outflow, and
the latter will be blown away. \citet{Owen2014f} estimated that this
radius to occur at approximately $10-20 \rplan$ for typical stellar wind
parameters. The corresponding time scales for flow variability are
once again in the range of a few hours to a few days.

For both cases considered above, the variability time-scale time
scales ranges from a few hours to a few days. The short end of the
range is roughly comparable to the transit duration and the long end
of the range is roughly comparable to the orbital period (for a Hot
Jupiter). Given this ordering of time scales, one would expect the
evaporative flow to vary from transit to transit. 

\subsubsection{Driven flow}

Driven flow can arise when the high energy flux of radiation received
by the planet displays time-variability. This paper has considered the
specific case where this variability results in a time-variable base
density that drives the flow. As discussed in
Section~\ref{sec:driven}, the response of the planetary evaporation
depends on whether the driving time-scale is shorter or longer than
the planet sound crossing time-scale ($\rplan/\cs\approx 3$ hours). If
the driving time is shorter than the crossing time, the flow cannot
respond to the rapid variations. As a result, the bulk quantities of
the flow are largely unchanged, as the fluid cannot globally respond
on a time-scale shorter than the flow crossing time. On the other
hand, if the driving time is longer than the crossing time, the fluid
does have time to respond; the density and velocity distributions will
thus vary on the driving time scale. In addition, the time-averaged
mass-loss rate follows an approximate power-law decline with
increasing variability time-scale.

Hot Jupiters have (at least) two obvious sources of flux variability
that can affect evaporation, including an eccentric orbit and/or
stellar flares. An eccentric orbit will lead to flux variability on a
time scale comparable to the orbital period, which is typically a few
days in these systems. In addition, stellar flares are observed to
vary on time scales of order a few hours to a few days \citep{guedel}.
As as result, we expect that most driven flows will have a response
firmly in the regime where the variability is longer than the planet
sound crossing time. The variability thus leads to a net decrease in
the time-averaged mass loss rate (see Figure \ref{fig:avg_mdot}). In
addition, with a variability on a time-scale of a few hours to days,
we expect the flow to vary from transit to transit. {\bc This indeed maybe the case for HD 189733b, where \citet{lecavelier2012} identified temporal variations in the Lyman-$\alpha$ absorption between two different epochs. As discussed above, the expected stellar and planetary magnetic field strengths mean spontaneous variability driven by the magnetic topology is unlikely. However, an X-ray flare was seen several hours before one of the transits with a factor 2-3 increase in flux. Our driven simulations show that variability of the high-energy flux on a time-scale of several hours ($T_{\rm var}\sim 1$), could lead to significant variability in the flow (see bottom panel of Figure~8). Thus, the X-ray flare could be responsible for the variability seen in the outflow of HD 189733b as hypothesised by \citep{lecavelier2012}.  }

 Targeted
observations of known eccentric planets, or co-eval observations in
the X-rays to detect flares, could thus be used to study driven
variability; {\bc this would be particularity useful in the case of the HD 189733 system as contemporaneous interactions between flares and outflow are already indicated.}

\subsection{Implications of observing variability}

Since the presence of any time variations in the flow is directly
linked to the strength of the planetary magnetic field, the results of
this paper provide an intriguing opportunity to place constraints on
the magnetic field strengths of exoplanets. For example, such a 
constraint could be found by producing a diagram similar to
Figure~\ref{fig:compare} and looking for the transition between
planets which show variability and those which do not.

Understanding the origin of inflated Hot Jupiters remains an unsolved
problem \citep[e.g.,][]{laughlin2011}. In one of the currently favoured
explanations, inflation arises from ohmic heating due to energy
dissipation from planetary winds interacting with the planetary
magnetic field \citep[e.g.,][]{batygin2010}. In this case, the
planetary magnetic field strength is a key parameter and theoretical
models tend to favour relatively large field strengths ($\bplan$
$\gtrsim$ 10~gauss) in order to account for the observed planetary
radii \citep[e.g.,][]{batygin2010,batygin2011,menou2012}. One should
keep in mind that ohmic dissipation occurs at much deeper layers in
the planetary atmosphere than the launch of outflows, as considered in
this paper. Nonetheless, both physical processes are tied to the
strength of the magnetic field strength at the planetary surface. As a
result, detections of flow variability consistent with magnetically
controlled outflows could be used to place constraints on the field
strength and hence level of ohmic dissipation.

\section{Conclusion}
\label{sec:conclude} 

This paper has considered magnetically controlled outflows from Hot
Jupiters in the regime where the magnetic field configurations do not
allow the flow to pass smoothly through the sonic point. In this regime, 
the resulting outflows are necessarily time-varying. Here we present a
summary of our main results, a brief discussion of their implications,
and some recommendations for future work.

\subsection{Summary of Results}   

[1] We have developed a scheme to consider the flow along a particular
magnetic field line, where the flow geometry (including, e.g., the
divergence operators) are taken into account using coordinate systems
developed in earlier work \citep{Adams2011}. The numerical results
show that a critical value of $\betacr$ exists, where $\betastar$ is
the ratio of the stellar field strength to the planetary field
strength, where both are evaluated at the planetary surface.  Systems
with larger $\betastar>\betacr$ have open magnetic field lines, and
hence open streamlines, for which the flow cannot pass smoothly
through the sonic point. For these cases, steady-state solutions do
not exist and the flow must be time variable. Moreover, the critical
value $\betacr$ found numerically (Section \ref{sec:nonsteady}) is in
good agreement with that derived in the previous analytic treatment of
the problem \citep{Adams2011}.

[2] For the case of steady-state forcing (due UV heating from the
star), we have found the corresponding time-dependent flow solutions.
These solutions exhibit quasi-periodic behaviour (see Figure
\ref{fig:mdot_boundary}). The time scale (essentially the period) of
these variations depends on the location of the outer boundary (see
Figure \ref{fig:timescale_vs_boundary}). This dependence on the domain
size arises because the flow remains subsonic, so that information can
propagate inward from the outer boundary and affect the flow. In real
planetary systems, the location of the effective outer boundary for
the planetary outflow is determined by the interface between the
sphere of influence of the planet and that of the star 
\citep{Owen2014f,Adams2011}. In dimensionless units, the time scale
of the variations is given by $T_{\rm evolve}\sim\rout^2/b$, 
which corresponds to the sound-crossing time of the scale height of
the flow evaluated at the outer boundary (see equation
[\ref{scaleheight}]).

[3] To elucidate the physics of these time-varying, subsonic,
magnetically-controlled flows, we have also considered systems where
the driving (heating) function is time-varying. This complication is
implemented by forcing the density at the base of the flow to be time
dependent, where we consider both sinusoidal and impulse
approximations. For the sinusoidal case, the driving function is
determined by an amplitude $A$ and a time scale $T_{\rm var}$. The
time-averaged mass loss rate depends on the ratio of the driving time
scale $T_{\rm var}$ to the sound crossing time of the planet. For
short driving time scales, $T_{\rm var} \ll 1$, the sinusoidal
variations are effectively averaged out, and the time-averaged mass
loss rate is nearly constant as a function of $T_{\rm var}$. In the
opposite limit, where the driving time scale is long, $T_{\rm var} \gg
1$, the flow has time to adjust and is only driven for a fraction of
the time; this complication leads to the time-averaged mass loss rate
decreasing with the driving time. In the cross-over regime, where
$T_{\rm var} \sim 1$, the mass loss rate displays a moderate resonant
enhancement (see Figure \ref{fig:avg_mdot}).

[4] The case of pulsed driving leads to similar behaviour. The
time-averaged mass-loss rate is nearly constant as a function of the
pulse separation time for $t_{\rm var} \lesssim1$, and decreases with
increasing $t_{\rm var}$ for larger values (see Figure 
\ref{fig:avg_mdot_pulse}). 

[5] The time variations in the flow are potentially observable, and
the considerations of this work can predict the relevant time scales
(which vary from a few hours to a few days).  For the simplest case
where the heating is steady, but the flow cannot pass smoothly through
the sonic point, the outflow displays quasi-periodic behaviour with a
time scale that depends on the effective size $\rout$ of the outflow
cavity (equation [\ref{eqn:tphysical}]). This size scale is set by the
outer boundary of the sphere of influence of the planet and depends on
the magnetic field strength of the planet, as well as the stellar
outflow and magnetic field parameters (this regime is determined by
the size of the dimensionless fields defined in \citealt{Owen2014f}).
With enough data, this line of inquiry can be used to constrain the
magnetic field strength on the planet by measuring the time scales.

[6] For the existing sample of close-in exoplanets, and for typically
expected values of the magnetic field strengths, planetary outflows
are almost always magnetically controlled by a safe margin (see also
\citealt{Adams2011,Owen2014f}).  Most planets orbiting older
(main-sequence) stars are likely to achieve steady-state transonic
outflows, but some fraction of evaporating planets are predicted to
display time-dependence (see Figure \ref{fig:compare}). For younger
(pre-main-sequence) stars, where stellar magnetic field strengths
$B_\ast \sim 1$~kG, a much larger fraction of planetary systems are
likely to experience time-variations in their outflows of the type
considered in this paper.

\subsection{Discussion and Future Work} 

The problem of planet evaporation separates into two regimes. Physical
processes in the inner region near the planet determine the launch of
the outflow from the planetary surface. In the present setting, the
geometry of the flow in this inner region is controlled by planetary
magnetic fields; the stellar background field also affects the flow
geometry and can prevent sonic transitions, where this complication
leads to time variability.  The outflow must eventually leave the
sphere of influence of the planet and enter into an outer region that
is dominated by external factors, including the stellar wind and the
stellar magnetic field. The nature of the flow in this outer region
will depend on the ratio of the ram pressure of the stellar wind to
the pressure of the stellar magnetic field, where both are evaluated
near the location of the planet. The outer boundary of the inner region
(equivalently, the inner boundary of the outer region) sets an
important length scale in the problem, where this scale is represented
by the outer boundary $\rout$ of the simulations in this paper. 
For example, this boundary radius determines the time scale for the
quasi-period variations seen in the outflow (where $t \sim \rout^2$). 
In order to investigate exactly how this variability will manifest
itself in observations, particularly Lyman $\alpha$ studies, one will
need to couple our simulations to those which determine the large scale
flow and the interaction between the star and planetary flows, such as
those recently presented by \citet{matsakos2015}.

Although this paper has made progress toward understanding time
varying outflows from exoplanets, a great deal of additional work
remains to be done.  We have carried out these simulations for one
streamline (following a single magnetic field line) at a time. The
global magnetic field structure will determine the types of
streamlines that are present, so that a more global treatment should
be carried out. In addition, this work has considered the simplest
case where the planetary magnetic field is a dipole, and the
contribution from the star can be modelled as a constant field (in the
same direction as the planetary dipole). Many different magnetic field
configurations are possible and should be explored in the future.

\section*{Acknowledgments} 

We thank the anonymous referee for comments which improved the manuscript. We would like to thank Marcelo Alvarez for useful discussions. JEO is
supported by NASA through Hubble Fellowship grant HST-HF2-51346.001-A,
awarded by the Space Telescope Science Institute (which is operated
for NASA under contract NAS 5-26555). FCA is acknowledges support from
the Michigan Center for Theoretical Physics. This collaboration was
facilitated by the International Summer Institute for Modeling in
Astrophysics, which was hosted in 2014 by the Canadian Institute for
Theoretical Astrophysics and the University of Toronto.

\bibliographystyle{mn2e}
\bibliography{library}

\begin{appendix}

\section{Tensor Artificial Viscosity}
\label{sec:TV}

It is necessary to use a full tensor implementation of the artificial
viscosity rather than a von Neumann \& Richtmyer approach
\citep{VonNeumann1950}. Thus, the artificial viscosity is updated
throught the following sub-step: 
\begin{equation}
\frac{\partial u}{\partial t}=-\frac{\nabla\cdot{\bf Q}}{\rho}
\end{equation}
where $Q$ is a trace-less, diagonal artificial viscosity tensor. 
In our coordinate system $(\nabla\cdot{\bf Q})_p$ is given by:
\begin{eqnarray}
(\nabla\cdot{\bf Q})_p&=&\frac{1}{h_qh_\phi}\left[\frac{\partial}{\partial p}\left(\frac{h_qh_\phi}{h_p}Q_{pp}\right)\right] -\frac{Q_{pp}}{h_p^2}\frac{\partial h_p}{\partial p}\nonumber \\&-&\frac{Q_{qq}}{h_ph_q}\frac{\partial h_q}{\partial p}-\frac{Q_{\phi\phi}}{h_ph_\phi}\frac{\partial h_\phi}{\partial p}
\end{eqnarray}
where
\begin{eqnarray}
Q_{pp} &=& \ell^2\rho\left(\nabla\cdot{\bf u}\right)\left[\left(\nabla{\bf u}\right)_{pp}-\frac{1}{3}\nabla\cdot{\bf u}\right]\\
Q_{qq} &=& \ell^2\rho\left(\nabla\cdot{\bf u}\right)\left[\left(\nabla{\bf u}\right)_{qq}-\frac{1}{3}\nabla\cdot{\bf u}\right]\\
Q_{\phi\phi} &=& \ell^2\rho\left(\nabla\cdot{\bf u}\right)\left[\left(\nabla{\bf u}\right)_{\phi\phi}-\frac{1}{3}\nabla\cdot{\bf u}\right]
\end{eqnarray}
where $\ell$ is a length scale taken to be $C_v h_p \Delta p$ where $C_v$ is an order unity constant that sets the number of grid cells over which a discontinuity is smoothed. The covariant derivatives of the velocity are given by:
\begin{eqnarray}
\left(\nabla{\bf u}\right)_{pp}&=&\frac{1}{h_p}\frac{\partial u}{\partial p}\\
\left(\nabla{\bf u}\right)_{qq}&=&\frac{u_p}{h_ph_q}\frac{\partial h_q}{\partial p}\\
\left(\nabla{\bf u}\right)_{\phi\phi}&=&\frac{u_p}{h_ph_\phi}\frac{\partial h_\phi}{\partial p}
\end{eqnarray}
In order to calculate the partial derivatives of the scale factors
with respect to $p$ one needs to make use of the following results:
\begin{eqnarray}
\frac{\partial \xi}{\partial p}&=&
\frac{\left(\betastar+2\xi^{-3}\right)\cos\theta}
{\left(\betastar+2\xi^{-3}\right)^2\cos^2\theta+\left(\betastar-\xi^{-3}\right)^2\sin^2\theta}\\
\frac{\partial \theta}{\partial p}&=&
-\frac{\xi^{-1}\left(\betastar-\xi^{-3}\right)
\sin\theta}{\left(\betastar+2\xi^{-3}\right)^2\cos^2\theta+\left(\betastar-\xi^{-3}\right)^2\sin^2\theta}
\end{eqnarray}
where we note the formula for $\partial\theta/\partial p$ in
\citet{Adams2011} contains a typo (Equation~47). Following
\citet{Adams2011} and defining the ancillary functions: 
\begin{eqnarray}
f &=& \betastar+2\xi^{-3}\\
g &=& \betastar-\xi^{-3}\\
H &=& f^2\cos^2\theta+g^2\sin^2\theta 
\end{eqnarray}
we obtain the following expressions for the derivatives of the scale
factors with respect to the coordinate $p$ 
\begin{eqnarray}
\frac{\partial h_p}{\partial p} & =& \frac{g\sin^2\theta\cos\theta\left(g^2-f^2\right)\xi^{-1}}{H^{5/2}} \nonumber \\
& &-\frac{3f\xi^{-4}\left(g^2\sin^2\theta-2f^2\cos^2\theta\right)\cos\theta}{H^{5/2}}\\
\frac{\partial h_q}{\partial p} & = & \frac{gf^{1/2}\sin^2\theta\cos\theta\left(g^2-f^2\right)\xi^{-1}}{H^{5/2}}-\frac{3f^{1/2}\cos\theta}{\xi^4H^{3/2}} \nonumber \\
& &-\frac{3f^{3/2}\xi^{-4}\left(g^2\sin^2\theta-2f^2\cos^2\theta\right)\cos\theta}{H^{5/2}}\\ 
\frac{\partial h_\phi}{\partial p} & = & \frac{3\sin\theta\cos\theta}{H\xi^{3}}
\end{eqnarray}
In passing, we note that in our numerical code we find that
numerically differencing the scale factors on our staggered mesh does
not result in a drop in accuracy. This is useful in geometries (for
example fields with many order multiples) where obtaining analytic
expressions for the partial derivatives of the scale factors may prove
time-consuming.

\section{Outflow Boundary Conditions}
\label{sec:BC}

In the regime where the outflow cannot pass smoothly through a sonic
transition, we anticipate that the velocity at the outer boundary is
not, in general, super-sonic. As a result, the ``standard'' outflow
boundary conditions, such as those commonly used in astrophysical
hydrodynamics codes (e.g., {\sc zeus} \citealt{Stone1992}), are not
applicable. In order to prevent spurious reflections at the outer
boundary, we must implement boundary conditions that only allow
outward travelling waves and suppress inward travelling waves. In
order to enforce this behaviour, we must solve the equations in
characteristic form at the outer boundary \citep{Thompson1987}.
We start by writing the primitive equations (Equations ~[\ref{eqn:rho}] 
and [\ref{eqn:u})] in matrix form: 
\begin{equation}
\frac{\partial {\bf U}}{\partial t}+
\frac{{\bf A}}{h_p}\frac{\partial {\bf U}}{\partial p}
+{\bf S}={\bf 0}\,.
\end{equation}
Then we find:
\begin{equation}
{\bf U}=\left(\begin{array}{c}
\rho\\
u\end{array}\right)\;\;{\bf A}=\left(\begin{array}{cc}
u & \rho \\
{\cs^2}/{\rho} & u\end{array}\right)
\end{equation}
and 
\begin{equation}
{\bf S}=\left(\begin{array}{c}
\frac{\rho u}{h_ph_qh_\phi}\frac{\partial}{\partial p}\left(h_qh_\phi\right) \\
\frac{1}{h_p}\frac{\partial \psi}{\partial p}\end{array}\right)
\end{equation}
The eigenvalues of ${\bf A}$ are $\lambda_1=u-c$ and $\lambda_2=u+c$ 
and the corresponding eigenvectors are:
\begin{equation}
{\bf e}^{L}_1=\left(-\cs,\rho\right) \;\;{\bf e}^{L}_2=\left(\cs,\rho\right)
\end{equation}
The characteristic equations are then given by:
\begin{equation}
{\bf e}^L_i\frac{\partial {\bf U}}{\partial t}+\lambda_i{\bf e}^L_i\frac{1}{h_p}\frac{\partial {\bf U}}{\partial p}+{\bf e}^L_i{\bf S}={\bf 0}
\end{equation}
following \citet{Thompson1987} and writing 
$\lambda_i{\bf e}^L_i/h_p\partial {\bf U}/\partial p$ as the operator
$\mathcal{L}_i$ and noting for outflow boundary conditions with no
spurious reflections we set $\mathcal{L}_i=0$ for any incoming
waves. Namely if $(u-c)/h_p<0$ at the outer boundary then
$\mathcal{L}_1=0$ or if $(u+c)/h_p<0$ at the outer boundary then
$\mathcal{L}_2=0$. Otherwise: 
\begin{eqnarray}
\mathcal{L}_1&=&\left(u-c\right)\left(-\frac{\cs}{h_p}
\frac{\partial \rho}{\partial p}+\frac{\rho}{h_p}\frac{\partial u}{\partial p}\right)\\
\mathcal{L}_2&=&\left(u+c\right)\left(\frac{\cs}{h_p}
\frac{\partial \rho}{\partial p}+\frac{\rho}{h_p}\frac{\partial u}{\partial p}\right)
\end{eqnarray}
We can find the evolution equations for the density and velocity 
in the boundary cells using Equations~(B5--B7), which leads to 
the forms 
\begin{eqnarray}
\frac{\partial \rho}{\partial t}-\frac{\mathcal{L}_1-\mathcal{L}_2}{2\cs}+
\frac{\rho u}{h_ph_qh_\phi}\frac{\partial}{\partial p}\left(h_qh_\phi\right)&=&0\,,\\
\frac{\partial u}{\partial t}+\frac{\mathcal{L}_1+\mathcal{L}_2}{2\rho}+
\frac{1}{h_p}\frac{\partial \Psi}{\partial p}&=&0\,.
\end{eqnarray}
As a result, Equations (B8--B9) can be integrated explicitly to find
the density and velocity in the boundary cells.

\section{Code Tests}
\label{sec:code_tests}

\begin{figure}
\centering
\includegraphics[width=\columnwidth]{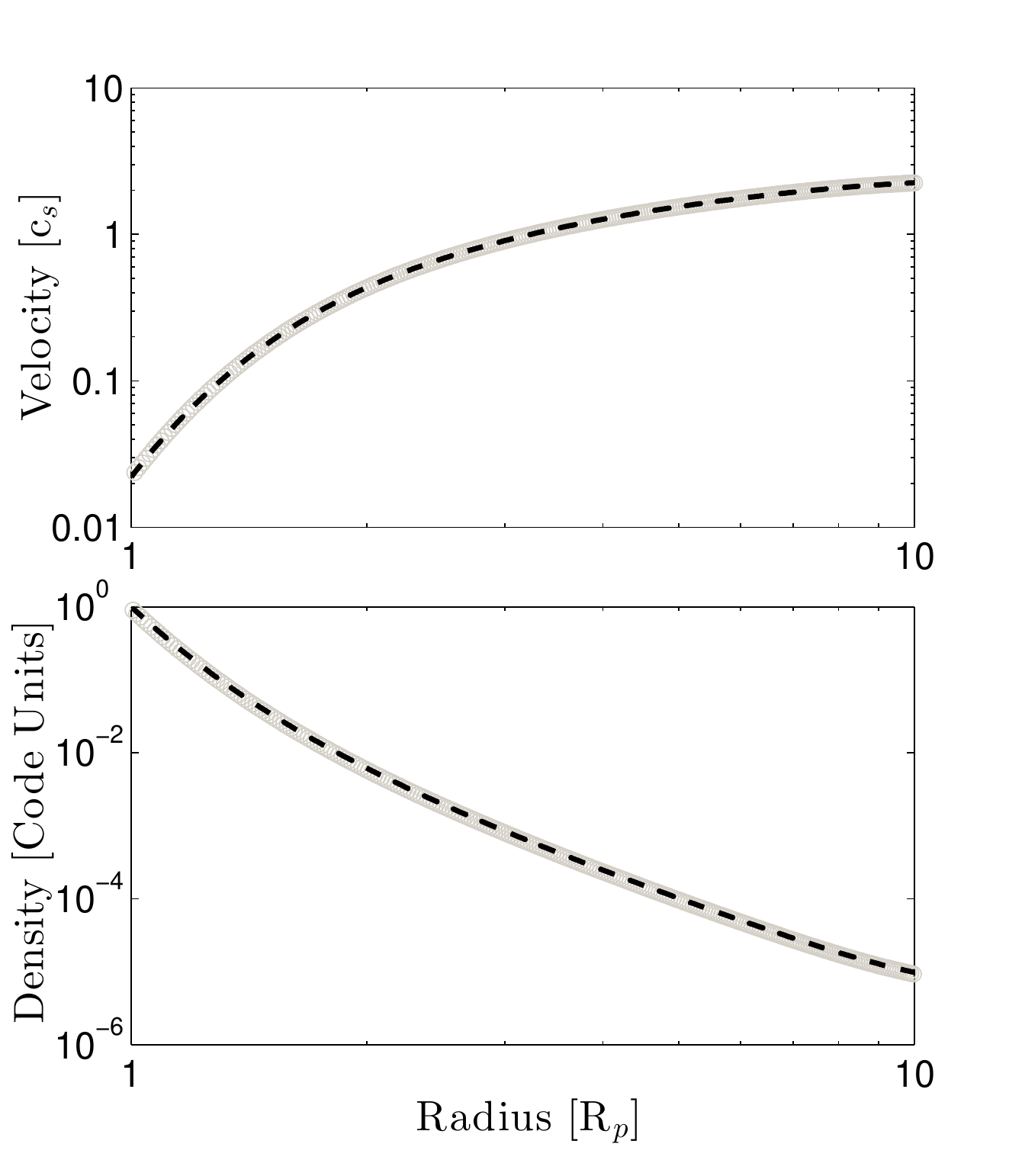}
\caption{Comparison of the flow profiles obtained from our simulations
(grey points) and analtyic solution (dashed line). The relative error 
between the two curves is approximately 0.003. }
\label{fig:follow_b_test}
\end{figure}

In order to determine our code is working as expected we perform a
simulation that matches onto analytic outflow solutions for
magnetically controlled winds from \citet{Adams2011}. It is thus easy
to derive a close form solution for the transonic outflow case
following \citet{Cranmer2004} to obtain: 
\begin{equation}
u^2=-W\left\{-\left(\frac{\lambda_s^2H}{q^2}\right)
\exp\left[2\left(b-\frac{b}{\xi}-\frac{\lambda^2_sH_1}{2q^2}
\right)\right]\right\}\,,
\label{eqn:analytic_transonic}
\end{equation}
where $W$ is the Lambert W function and $\lambda_s$ is the mass-flux
of the transonic solution given by Equation~(51) of \citet{Adams2011}.
We have performed several simulations to check if we can numerically
reproduce the analytic solution given in
Equation~\ref{eqn:analytic_transonic}.  The simulations proceed as
follows: The grid initialised with zero velocity and low density
$10^{-5}\rhobase$, apart from a set of inner ghost cells set to density 
$\rhobase$. Simulations are run with units of $G=\cs=\rhobase=\rplan=1$ 
from $\xi=1$ to 10; the sonic point lies at $\xi\approx3.23$, where the 
starting angle is 0.3 radians and $\betastar=10^{-3}$. We use a grid
with 256 cells and run the simulation for $t=200$ time units
(corresponding to $\sim$ 20 sound crossing times). The velocity and
density profile found from the simulation (grey points) are compared
to the analytic solution (dashed line) in 
Figure~\ref{fig:follow_b_test}. The relative error between the
numerical and analytic solution is $\sim 3\times 10^{-3}$ (which 
is smaller than the width of the lines in the figure). We thus 
conclude that our simulations are behaving as expected.

\end{appendix}

\end{document}